\documentclass[manuscript]{aastex631}
\usepackage{CJK}
\usepackage{graphics}
\usepackage{graphicx} 
\usepackage{booktabs}

\begin{document}
\begin{CJK*}{UTF8}{gbsn}
\title{Real-time Light Curve Classification Framework for the Wide Field Survey Telescope Using Modified Semi-supervised Variational Auto-Encoder}

\correspondingauthor{Yongling Tang, Lulu Fan}
\email{tangyongling@mail.ustc.edu.cn, llfan@ustc.edu.cn}

\author[0009-0003-0819-8357]{Yongling Tang (唐永灵)}
\affiliation{Department of Astronomy,
University of Science and Technology of China, Hefei 230026, China}
\affiliation{School of Astronomy and Space Science, University of Science and Technology of China, Hefei 230026, China}

\author[0000-0003-4200-4432]{Lulu Fan (范璐璐)}
\affiliation{Department of Astronomy,
University of Science and Technology of China, Hefei 230026, China}
\affiliation{School of Astronomy and Space Science, University of Science and Technology of China, Hefei 230026, China}
\affiliation{Deep Space Exploration Laboratory, Hefei 230088, China} 

\author[0000-0002-3105-3821]{Zhen Wan (宛振)}
\affiliation{Department of Astronomy,
University of Science and Technology of China, Hefei 230026, China}
\affiliation{School of Astronomy and Space Science, University of Science and Technology of China, Hefei 230026, China}

\author[0009-0008-4858-1410]{Yating Liu (刘雅婷)}
\affiliation{School of Artificial Intelligence and Data Science, University of Science and Technology of China, Hefei 230026, China}
\affiliation{Shanghai AI Laboratory, Shanghai 200232, China}

\author[0009-0002-1449-5174]{Yan Lu (陆岩)}
\affiliation{Shanghai AI Laboratory, Shanghai 200232, China}

\begin{abstract}

Modern time-domain astronomy will benefit from the vast data collected by survey telescopes. The 2.5-meter Wide Field Survey Telescope (WFST), with its powerful capabilities, is promising to make significant contributions in the era of large sky surveys. To harness the full potential of the enormous amount of unlabeled light curve data that the WFST will collect, we have developed a semi-supervised light curve classification framework. This framework showcases several unique features: First, it is optimized for classifying events based on the early phase of the light curve (three days after trigger), which can help identify interesting events early and enable efficient follow-up observations. Second, the semi-supervised nature of our framework allows it to leverage valuable information from large volumes of unlabeled data, potentially bridging the gap between simulations and real observations and achieving better generalization in practical scenarios. Compared to the commonly used Recurrent Neural Network (RNN) models, our framework has shown a 5.59\% improvement in accuracy for early classification tasks, as well as improvements in precision and recall in almost all subclasses. Moreover, our approach provides a reconstructed light curve, along with a compact latent representation, offering a different perspective that can be used for further downstream tasks beyond classification. The code and model weights used in this work are maintained and publicly available in \href{https://github.com/vetkimi/WFST-light-curve-classifier}{GitHub repository}.

\end{abstract}


\section{Introduction}\label{sec:intro}

With the advance of modern wide field survey infrastructures such as the Catalina Real-Time Transient Survey \citep[CRTS;][]{djorgovski2011catalina}, the Asteroid Terrestrial-impact Last Alert System\citep[ATLAS;][]{jedicke2012atlas}, the Dark Energy Survey \citep[DES;][]{dark2016dark}, the Panoramic Survey Telescope and Rapid Response System \citep[Pan-STARRS;][]{kaiser2010pan}, the Zwicky Transient Facility \citep[ZTF;][]{bellm2018zwicky}, and the Legacy Survey of Space and Time \citep[LSST; ][]{ivezic2019lsst}, astronomers can explore the changing sky with greater breadth and depth. The drilled search for the changing sky is vital for astronomy discoveries, from studying the progenitor systems of transients to cosmological distance measurements and discovering new phenomena \citep[][]{riess1998observational, villar2018superluminous, pursiainen2018rapidly}. However, with this opportunity comes various kinds of data processing challenge. One of these challenges is to facilitate early-time classification of the light curve among the vast data. An early-time classification process will lead to more rapid and effective follow-up observations and ultimately provide a higher chance of scientific discovery. In the era of large-sky surveys, the 2.5-meter Wide Field Survey Telescope (WFST) is about to dedicate itself to the front line of exploring the dynamic universe \citep{wang2023sciences}. Due to the multiple time domain targets (see Section~\ref{sec:WFST}) of WFST and the limitation of follow-up observation resources, a specifically designed light curve classification tool for WFST is essential to unleash its potential.

Due to the need for automated classification tools for the current and upcoming wide-field surveys, research on such topics has been widely conducted. Due to the quantity and complexity of the data, traditional methods such as the fitting of templates \citep{sako2007sloan} cannot fully use the potential value of all the data. Encouraged by the success of machine learning in recent years, researchers tend to find machine learning solutions to astronomical data mining problems \citep[][]{baron2019machine, yu2021survey}. Generally, light curve classification techniques can be divided into two categories: 1) translating light curves into hand-made features, which are then fed into machine learning algorithms such as random forests \citep{sanchez2021alert, reis2018probabilistic, narayan2018machine}; and 2) instead of adopting handcrafted features, using a deep neural network algorithm to extract useful features from the data (also known as representation learning). Examples of these techniques in the second category include convolutional neural networks \citep[][]{aguirre2019deep}, recurrent neural networks \citep[][]{naul2018recurrent}, encoder-decoder architecture \citep[][]{pasquet2019pelican, jamal2020neural}, stacked neural networks \citep[][]{chaini2020astronomical} and Transformer \citep[][]{pimentel2022deep, donoso2023astromer,allam2023tiny, allam2024paying}. There are also some hybrid approaches that combine deep learning with traditional machine learning techniques. In these approaches, the final dense classification layer of the neural network model is replaced with more conventional methods, such as Random Forest. This integration allows to benefit from the strengths of both deep learning (for feature extraction) and traditional machine learning (for classification), potentially improving the robustness and interpretability of the model \citep{boone2021parsnip, villar2020superraenn}.

Although considerable effort has been devoted to the field of automatic light curve classification, two issues still demand further attention. Firstly, the majority of classification approaches focus on full light curve classification. Secondly, most of the classification methods do not make the most use of the unlabeled data collected by the survey telescope. Regarding the first issue, full light curve classification necessitates that each light curve sample include sufficient observations to cover the primary evolution of transient events, thereby missing the opportunity to classify light curves and initiate follow-up observations during their early stages. This rapid follow-up scheme is crucial for the timely investigation of fast-evolving transient phenomena, which rapidly fade away. The second concern is especially crucial for new survey telescopes. Although simulations can create useful training datasets, differences between simulated and real data may still exist. Some previous work has tried to deal with these two problems. For the first problem, the recurrent neural network-based classifier \citep[][]{muthukrishna2019rapid, moller2020supernnova} is introduced to extract information from the limited early observation. Efforts have also been made to convert early-time light curves into heat maps so that well-established image classification algorithms (convolutional neural networks) can be applied \citep{qu2022photometric}. These works achieved decent early light curve classification results in simulation data. However, the connection between early observation and the later evolution of the light curve is not well established, which could offer additional information for the early classification task. For the second problem, various data augmentation methods have been used to augment the bias training data. Unsupervised learning techniques are also introduced to ease the bias training data problem by leveraging the unlabeled data, as these massive unlabeled data contain more comprehensive information, as well as the characters of the telescope and the new unexplored categories \citep{richards2012semi,villar2020superraenn,pasquet2019pelican}. However, the unsupervised methodology derives a low-dimensional representation of the unlabeled data, which may not be inherently advantageous for classification tasks.

In this work, we develop a novel framework to address the early light curve classification problem. This framework incorporates an unsupervised autoencoder that facilitates establishing a connection between early light curve information and its later evolution. It also enables the extraction of information from large volumes of unlabeled data, therefore offering a more effective solution to the above-mentioned limitations. This paper is organized as follows. In Section 2, we introduce the WFST project and the simulated datasets. In Section 3, we describe the methods employed in this work and detail how we construct our classifier framework. In Section 4, we present the performance of our framework and discuss the results. Finally, we summarize our work in Section 5.

\section{Data}

\subsection{WFST}\label{sec:WFST}

The Wide Field Survey Telescope (WFST) is a photometric surveying facility jointly built by the University of Science and Technology of China (USTC) and the Purple Mountain Observatory (PMO). The telescope features a 2.5-meter diameter primary mirror with an active optics system. It is equipped with a mosaic CCD array comprising 0.73 gigapixels on the primary focal plane, providing a 6.5-square-degree field of view (FOV). Two key programs are planned for the WFST 6-year survey: the Wide-Field Survey (WFS) program and the Deep High-cadence u-band Survey (DHS) program. These survey modes, designed with varying depths, areas, and cadences, are in alignment with the primary scientific objectives of WFST. Each program will occupy approximately 45\% of the total observation time. The remaining 10\% of the observing time (about 1,300 hours over 6 years) will be allocated to smaller campaigns for specific purposes, such as capturing time-critical targets and intensively scanning certain skies of particular interest \citep{wang2023sciences}.

WFST will observe the sky in five optical bands ($u$, $g$, $r$, $i$, and $z$) with cadences ranging from hourly/daily in the DHS program to semiweekly in the WFS program. A nominal 30-second exposure in the WFS program will achieve a 5$\sigma$ detection limit of 22.31, 23.42, 22.95, 22.43 and 21.50 AB magnitudes in these bands during a photometric night \citep{Lei_2023}. In the DHS program, intra-night 90-second exposures reaching depths of 23 mag ($u$) and 24 mag ($g$), combined with target-of-opportunity follow-ups, will provide unique opportunities to explore the dynamic universe, including the electromagnetic counterparts of gravitational wave events \citep{Liuzhengyan_2023}, supernovae \citetext{SNe; \citealp{Humaokai_2023}}, tidal disruption events \citetext{TDEs; \citealp{LinZheyu_2022}}, and Active Galactic Nuclei \citetext{AGN; \citealp{Su_2024}}. The survey mode of WFST provides a unique opportunity to track transients right after their occurrences and to discover rare energetic explosive phenomena in the universe (e.g. early-phase supernovae, fast blue optical/ultraluminous transients, TDEs, kilonova, etc.). Timely multi-wavelength follow-up observations therefore become feasible. Detailed observations, including high-cadence photometry, time-resolved spectroscopy, and spectropolarimetry, shortly after the explosion provide insight into the progenitor systems that power the event and thus improve our understanding of the physical mechanism of these objects. Therefore, ensuring a short latency between a transient detection and its follow-up is an important scientific challenge.

In general, WFST is expected to detect a significant number of transients in the low-redshift universe and systematically investigate the variability of Galactic and extragalactic objects.  The final six-year co-added images, anticipated to reach $g \simeq 25.8$ mag in the WFS or 1.5 mag deeper in the DHS, will be of fundamental importance to general Galactic and extragalactic science. The co-added source catalogs are scheduled for annual public release. A real-time alert stream is currently being developed and will be accessible to the WFST collaboration \citep{cai2025WFST}. The highly uniform legacy surveys of the WFST will serve as an indispensable complement to those of the Vera C. Rubin Observatory Legacy Survey of Space and Time (LSST), which monitors the southern sky. Detailed information on WFST and its scientific goals can be found on its official website\footnote{\url{https://wfst.ustc.edu.cn/}}. 

\subsection{Simulation} 

In order to overcome the lack of representative data and comprehensively evaluate our new ideas, we adopt simulated data in this work. For a new survey project, the lack of sufficient labeled data always presented challenges in training a suitable classification algorithm. Although real data from other surveys might provide a solution, the different characteristics of different survey telescopes, such as filters, cadence, instrument characteristics, and observing conditions, as well as the absence of certain targets, hinder the direct use of these data. We choose the nine most common classes among the PLAsTiCC models, including AGN, TDE, kilonova (KN) and different supernovae variants such as Type Ia (SNIa), Type Ibc (SNIbc), Type II (SNII), SN1991bg-like objects (SNIa-91bg), peculiar supernovae (SNIax), and superluminous supernovae (SLSN). By utilizing the SNANA software package \citep{kessler2009snana}, we are able to generate simulated light curves that might reflect the true features of the survey. The SNANA code takes into account several variables that could affect the characteristics of the observed light curve in the simulation process. The PLAsTiCC model is rich in content on transient sources and has contributed significantly to the research on automatic classification of light curves \citep[][]{burhanudin2023pan,hlovzek2023results,abdelhadi2024picture,russeil2024rainbow}.

\begin{figure}[!htb]
  \centering
  \includegraphics[width=0.6\textwidth]{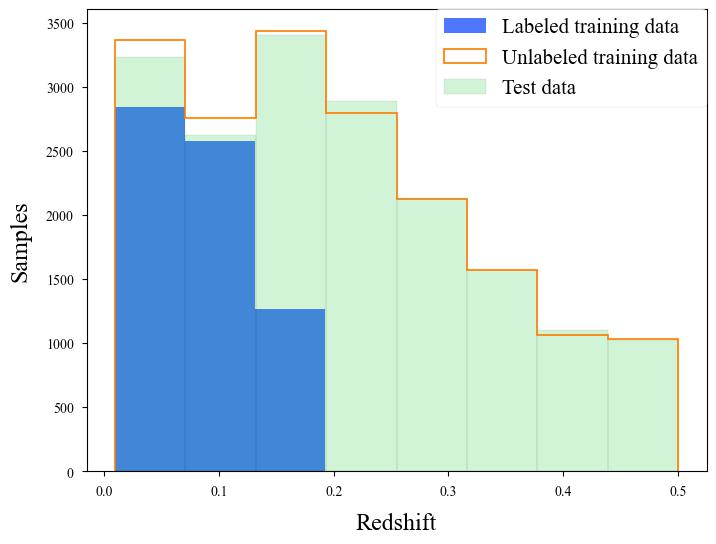}
  \caption{Redshift distributions of three datasets representing the following transient classes: KN, SLSN, SNIa, SNIa-91bg, SNIax, SNIbc, SNII and TDE.}
  \label{fig:figure-1}
\end{figure}

\begin{figure}[!htb]
  \centering
  \includegraphics[width=1.0\textwidth]{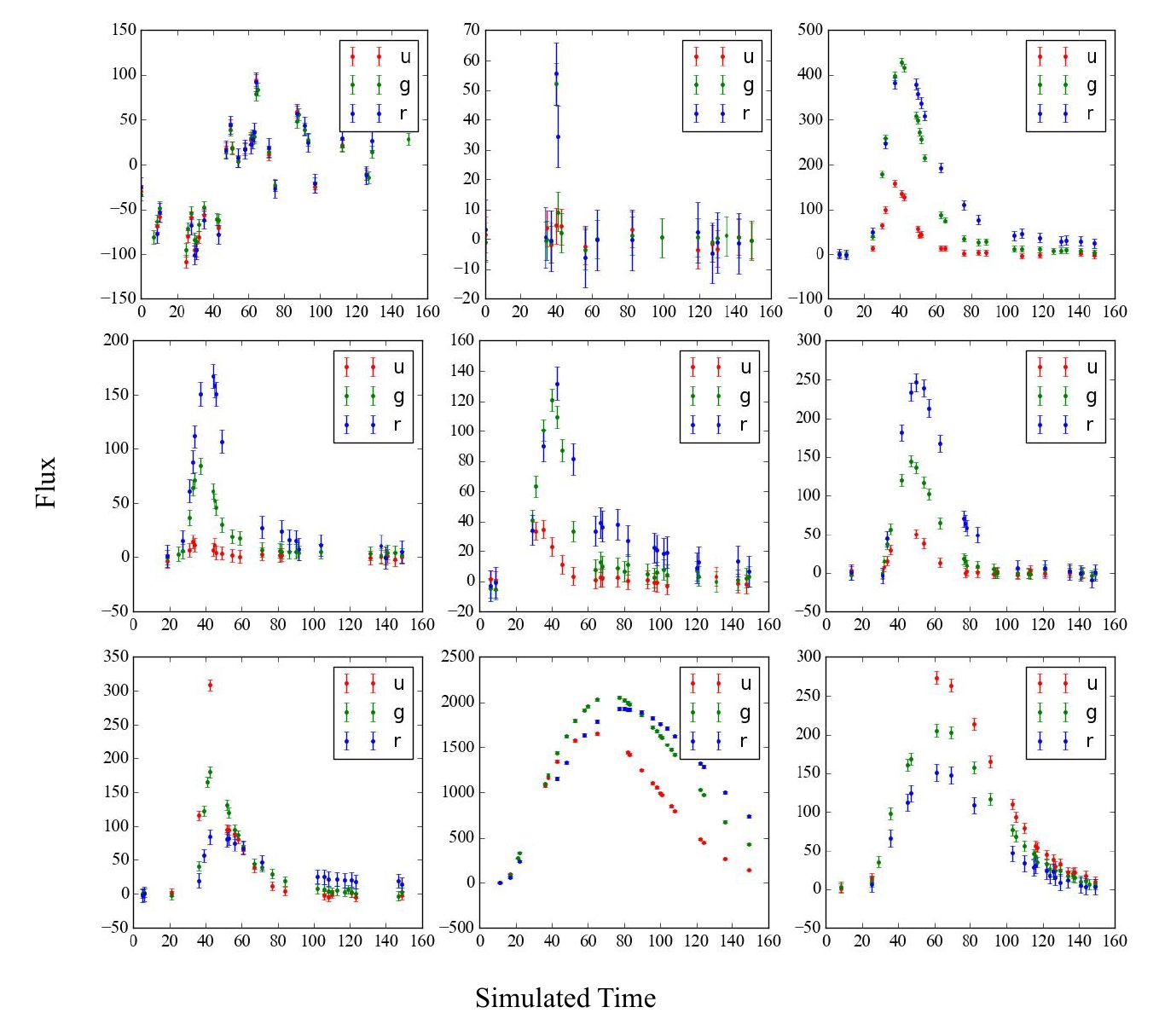}
  \caption{Illustrations of well sampled simulated light curves in expected WFST cadence. From left to right, the light curves represent the following classes: AGN, KN and SNIa (the first line); SNIa-91bg, SNIax and SNIbc (the second line); SNII, SLSN and TDE (the third line).}
  \label{fig:figure-2}
\end{figure}

We generated three datasets in total: a labeled training dataset, a test dataset, and an additional unlabeled training dataset. The ratio of these three datasets is 1:1:2 after preprocessing, as described in Section~\ref{sec:preprocess}. We deviate from the commonly used ratio of training to test sets to emphasize the importance of unlabeled data. Figure~\ref{fig:figure-1} demonstrates the different redshift distributions of three datasets. Table~\ref{tab:table_1} shows the number of instances in each class. Our simulation utilizes a variety of models from PLAsTiCC \citep{plasticc_modelers_2022_6672739}.We utilize a majority of the standard simulation parameters, including cosmological values, the correlation between simulation rate and redshift, and Galactic extinction, among others, as specified in the SNANA manual. Furthermore, we create our own `SIMLIB' file and `Survey definition' necessary for the SNANA software to produce light curves that resemble those of the WFST survey. Figure~\ref{fig:figure-2} shows some examples of the simulated light curves. The detailed preprocessing of these datasets is described in Section~\ref{sec:preprocess}.

\begin{table}
\caption{Total Number of Each Class}
    \centering
    \begin{tabular}{cccccccccc}
            & AGN & KN & SLSN & SNIa & SNIa-91bg & SNIax & SNIbc & SNII & TDE \\
      \toprule
      Before augmentation & 3000 & 3000 & 5000 & 6000 & 3708 & 2931 & 887 & 2908 & 5000\\
      \midrule
      After augmentation & 4000 & 4000 & 5000 & 6000 & 4278 & 3431 & 1387 & 3408 & 5000 \\
      \bottomrule
    \end{tabular}
    \label{tab:table_1}
\end{table}

In this study, we define an early classification as one performed three days after trigger, while a full classification occurs 90 days afterward. We ensure that each simulated light curve contains at least three measurements before the trigger, as they helped establish a baseline flux \citep{muthukrishna2019rapid}. We truncate the original simulated light curves to have 90 days of coverage (starting three days before the trigger), which we refer to as a ``full light curve". The definition of a ``trigger" for an alert event in the light curve is any observation that exceeds the 5$\sigma$ signal-to-noise (S/N) threshold of the WFST imaging difference tool \citep{hu2022image}. In the transient detection pipeline, imaging difference is the most widely used method to detect possible transient events. The imaging difference technique subtracts the reference image from a new image to detect variable sources. It has been proven to be effective, especially in crowded fields \citep[][]{tomaney1996expanding,bond2001real}.

\section{Methods}

Our methods include a Gaussian Process Regression (GP) for data preprocessing and a variational autoencoder-based semi-supervised framework for classification. In Sections~\ref{sec:GP} and \ref{sec:dl-methods}, we briefly introduce these techniques. Details about data preprocessing and the structure of the framework are provided in Section~\ref{sec:framework}.

\subsection{Gaussian Process Regression} \label{sec:GP}

Given the intricate nature of light curves, we utilize GP for both interpolation and augmentation of these datasets. These characteristics include: (i) temporal irregularities, (ii) data collected through multiple filters, (iii) streaming data formats, and (iv) uneven distributions among various types. These factors hinder the straightforward application of deep learning techniques, necessitating the steps of interpolation and data augmentation. GP offers a robust approach to these challenges, capable of effectively modeling time series data, such as light curves and spectral energy distributions (SEDs). By accounting for uncertainties inherent in the data, the GP is resilient against the influence of low-confidence outliers. In astronomical contexts, GP can generate reliable interpolated light curves even when dealing with sparse and noisy datasets, while also providing credible error estimates. For example, \citet{villar2020superraenn} utilized GP to interpolate missing values in light curves before entering them into a machine learning model. This methodology has demonstrated its efficacy in several classification tasks \citep{revsbech2018staccato,villar2020superraenn, hosenie2020imbalance, moller2020supernnova}.

GP serves as a probabilistic framework for modeling light curves, characterized by a mean function (typically set to zero) and a covariance function, or kernel. Following \citet{boone2019avocado}, we adopt the Mat$\acute{\rm e}$rn-3/2 kernel, which is more suitable than the smoother squared exponential kernel to capture sudden changes in light curves, such as those arising from transient phenomena. \citet{boone2019avocado} also demonstrated the advantage of employing a two-dimensional GP, combining kernels for both time and wavelength to capture interband correlations. This approach enhances the ability to infer light curves, particularly when data is sparse. Unlike earlier studies, which often modeled each band separately, the multidimensional kernel facilitates the simultaneous modeling of all bands. Our implementation utilizes the \texttt{George} package \citep{2015ITPAM..38..252A} for GP modeling, with optimization carried out through the \texttt{Scikit-learn} library \citep{scikit-learn}. The resultant models offer nonparametric descriptions of light curves and their associated uncertainties, using crossband information to refine predictions in undersampled bands.

\subsection{Deep Learning Algorithms for the Classification Framework}\label{sec:dl-methods}

The deep learning method has several advantages over traditional machine learning methods. First,the fast speed of modern GPUs makes them suitable for real-time processing of enormous data streams from survey telescopes. Second, deep learning does not require handcraft features, thus avoiding the artificial bias that can arise from such features. In many cases, the features learned by deep learning models are comparable or even more effective than the features made by human experts. Our deep learning model is implemented using \texttt{Keras} \citep{chollet2015keras}.

\subsubsection{Recurrent Neural Network}

Recurrent neural networks (RNNs) serve as a robust algorithm for handling time series data. They process the data incrementally, retaining a state that captures the information from earlier time steps. This architecture allows the network to see the connection between current and previous time steps in sequential order. However, due to the optimization algorithm, the simplest version of RNNs suffers from the problem of gradient vanishing, which hampers the learning process. To address this issue, researchers have designed variations of RNNs, such as LSTM \citep[long-short-term memory;][]{hochreiter1997long} and GRU \citep[gated recurrent unit;][]{cho2014properties}. In astronomy, these recurrent neural network architectures exhibit high performance in classifying light curves compared to other machine learning architectures \citep{jamal2020neural}. Compared to the LSTM model, the GRU has fewer parameters, simplifies the structure of the model, and improves efficiency. The GRU model introduces the gating mechanisms at each time step to better manage memory and information flow. As a result, GRU can effectively capture and transfer long-term dependencies while maintaining short-term information. Therefore, we used the GRU architecture in this work.

\subsubsection{Variational Auto-encoder}

Inspired by previous works \citep{pasquet2019pelican, villar2020superraenn, boone2021parsnip}, our framework is built on the basis of a Variational Auto-Encoder (VAE). The VAE module serves two key roles in our method. First, it learns a useful latent representation from unlabeled data. Second, it captures the relationship between the early light curve and its later evolution, which enhances our model's ability to classify the light curve in the early phase. 

Auto-encoders (AE) are a class of neural networks whose purpose is to encode input data into a low-dimensional latent space and then decode it back to the original form. In astronomy, autoencoder architectures have been utilized for various purposes, including feature learning in galaxy spectral energy distributions \citep{frontera2017unsupervised}, image de-noising \citep{ma2018radio}, dimensionality reduction \citep{portillo2020dimensionality}, and event classification \citep{villar2020superraenn}. Instead of simply compressing the raw data into a fixed encoding in the latent space, VAE converts the data into the parameters of a statistical distribution. By doing so, it ensures that any position in the latent space corresponds to a meaningful representation, allowing every point to be decoded into a meaningful output. Additionally, any two adjacent points in the latent space can be decoded into highly similar data in VAE latent space. 

\subsection{Semi-supervised Classification Framework}\label{sec:framework}

\subsubsection{Data Preprocess}\label{sec:preprocess}

To preprocess the light curve data, we follow the following steps. First, we select the light curves in the most widely used bands of WFST, namely $u$, $g$, and $r$, from the simulated data. Next, we augment the normalized full light curves by generating new samples using GP. Then, we generate a partial light curve by masking the values after a certain time step with zeros. To present consistently formatted data for the neural network, we interpolate and truncate both the partial light curve and the corresponding full light curve to ensure they have a 90-point length (including padded zero values) with a one-day interval starting three days before the trigger. Finally, we perform data normalization. The rationale behind these preprocessing steps is as follows. First, each data point in a light curve is updated according to the given WFST cadence. Second, our framework is designed to handle real-time ``partial light curves''. We define a light curve with the latest observation date less than 90 days (starting three days before the trigger) as a ``partial light curve" (as opposed to the ``full light curve" mentioned earlier). These preprocessing steps ensure that the data are consistent, normalized, and suitable for both real-time and full light curve analysis, facilitating the effective use of the neural network.
\begin{figure}[!htb]
  \centering
  \includegraphics[width=1\textwidth]{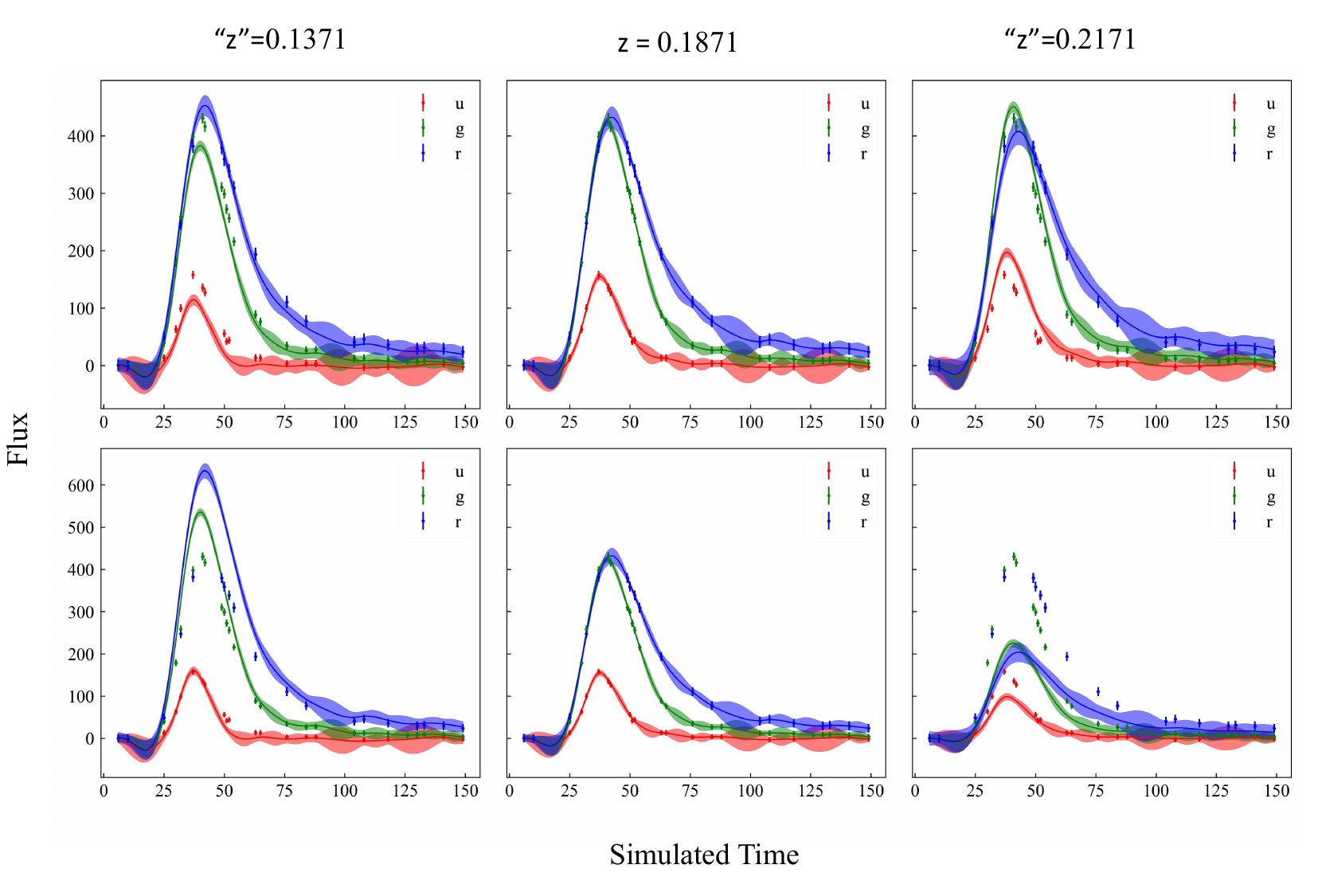}
  \caption{Illustration of GP Augmentation Applied to Simulated Light Curves. Each band displays the mean Gaussian Process (GP) flux prediction as a solid line, with a shaded contour representing one-standard-deviation uncertainty in the flux prediction. The middle column presents the initial fit. The left column shows the augmented data for a "low redshift" scenario (achieved by reducing the wavelength parameter in the GP model). The right column illustrates the augmented data for a "higher redshift" case (accomplished by increasing the wavelength parameter in the GP model). The top row demonstrates the augmentation solely by adjusting the wavelength parameter, whereas the bottom row reflects adjustments made after considering the distance difference. For ease of comparison, each panel includes error bars that align with the original light curve. (Note: Since these scenarios do not precisely reflect real redshifts, we use quotation marks around lower and higher ``z" to indicate this distinction from actual redshift values.) }
  \label{fig:figure-3}
\end{figure}

\begin{figure}[!htb]
  \centering
  \includegraphics[width=0.6\textwidth]{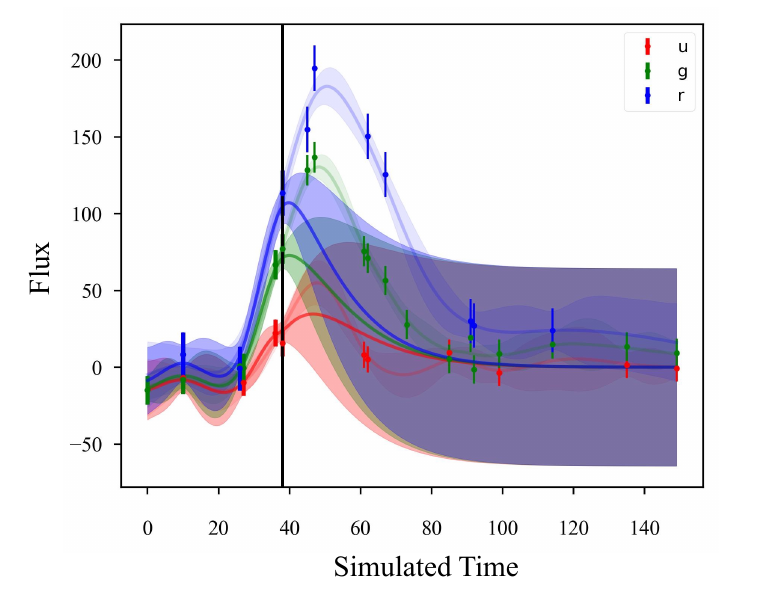}
  \caption{Example of GP real-time interpolation applied to simulated light curves. A solid vertical line divides the light curve into segments for early classification (left) and full classification (right). Each band's mean GP flux prediction is shown as a solid line, surrounded by a shaded contour indicating the one-standard-deviation uncertainty in the flux prediction. The GP model produces relatively consistent interpolation but performs poorly in forecasting future data points.}
  \label{fig:figure-4}
\end{figure}

Figure~\ref {fig:figure-3} shows an example of an augmented light curve. Similarly to \citet{boone2019avocado}, we modify the wavelength parameter (in the $u$, $g$, and $r$ bands) of the GP model for each light curve to enhance the data. Compared to the approach taken by \citet{boone2019avocado}, our work involves only a minor augmentation of the light curve. After the augmentation, the distribution of class frequencies within the labeled training dataset becomes approximately uniform. Figure~\ref{fig:figure-4} shows the real-time interpolation at different stages of the light curve. The majority of previous studies utilizing GP models employ the full light curve for interpolation objectives. However, in our analysis, only the data points available up to the latest observation in these light curves are utilized.\citet{villar2020anomaly} found that this limitation does not affect their results. Similarly, we also conclude that it will not affect our purpose, as GP can produce relatively consistent interpolations (see Figure~\ref{fig:figure-4}).
The final data set prepared for training and testing the neural network is organized into a tensor with shape (number of samples, time steps, number of features). The dimension of ``number of samples" corresponds to the total number of data instances.The ``time steps dimension" is set to 90, representing a 90-day observation window with a one-day interval between each data point. This window is aligned to start three days before the trigger event. The dimension of ``number of features" includes six features, comprising flux and flux error measurements in the three most commonly used bands ($u$, $g$, and $r$) of the WFST. Each band contributes two features: the flux value and its associated error. Given that the interpolated light curves are evenly distributed over time, we do not explicitly include the time as a separate feature.

We emphasize that proper data normalization is crucial, especially when a framework involves a reconstruction step. According to \citet{villar2020superraenn}, considerable variations in event scale can result in impaired performance of models, a conclusion that our experiments also supported. To ensure appropriate data normalization while preserving absolute brightness information, we applied the following normalization procedures, as defined in Equations~\ref{equ:equation-1} and \ref{equ:equation-2}. In the data normalization equations, the phase index $j$ indicates the latest observation date since the trigger. For example, the index $j=20$ represents the light curve sample with observations in a window of $20$ days (start three observations before trigger). In this case, all data points where $90\ge t>j$ are masked, and the network will not process these steps (the early light curve and the full light curve can be considered as two special cases of the partial light curve when $j=6, 90$ respectively). $F_{\mathrm{ij}}(t), Ferr_\mathrm{ij}(t)$ represent the flux and flux error measured in the sample index $\mathrm{i}$, the phase index $\mathrm{j}$ and the time step $t$. $\tilde{F}_{\mathrm{ij}}(t), \tilde{F} err_\mathrm{ij}(t)$ is the corresponding normalized flux and flux error. $F$ if the maximum flux among all training data.

\begin{equation}\label{normalized flux}
     \tilde{F}_{\mathrm{ij}}(t)= \frac{F_{\mathrm{ij}}(t)}{\max(F_{\mathrm{ij}}(t))} \times \frac{\log(\max(F_{\mathrm{ij}}(t)))}{\log(F)}
    \label{equ:equation-1}
\end{equation}

\begin{equation}\label{normalized flux err}
     \tilde{F}err_\mathrm{ij}(t) = \frac{Ferr_\mathrm{ij}(t)}{\max(F_\mathrm{ij}(t))} \times \frac{\log(\max(F_{\mathrm{ij}}(t))}{\log(F)}
    \label{equ:equation-2}
\end{equation}

\subsubsection{Semi-supervised Framework}

\begin{figure}[!htb]
  \centering
  \includegraphics[width=0.8\textwidth]{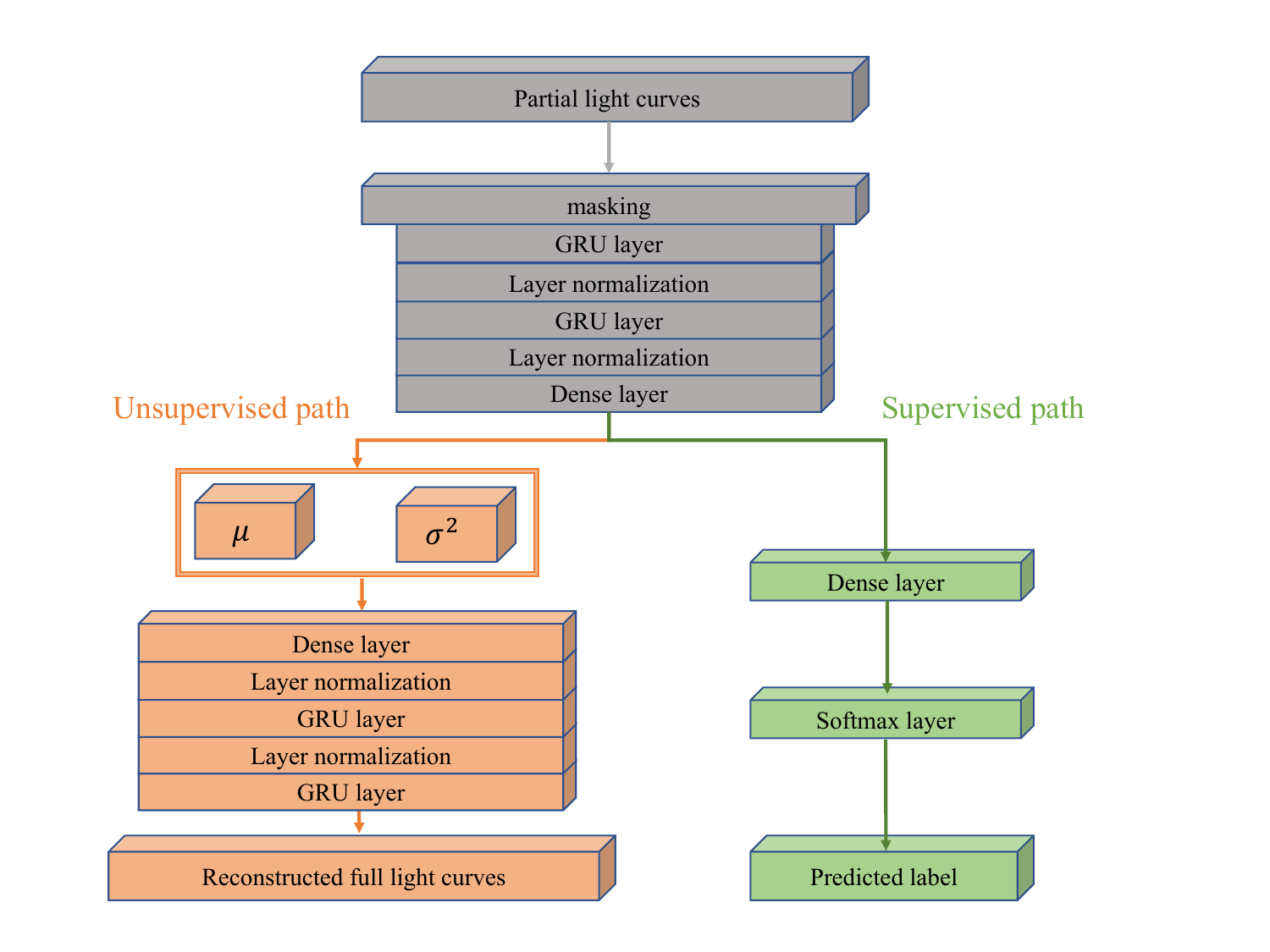}
  \caption{The semi-supervised classification architecture comprises two distinct pathways. The unsupervised component (denoted as the orange pathway) is purposed to discern the low-dimensional distribution inherent in the data. It functions by reconstructing the original full light curve from its partial counterpart, thereby establishing a linkage between the partial and full light curves. Meanwhile, the supervised component (depicted in green) is tasked with generating the predicted probability of classification for each class. Both pathways are concurrently trained and they utilize a shared encoder module, represented in grey.} 
  \label{fig:figure-5}
\end{figure}

The architecture of our framework is shown in Figure~\ref{fig:figure-5}. The architecture consists of two distinct paths: an unsupervised auto-encoder path and a supervised classification path. The Unsupervised Auto-Encoder Path is implemented as a VAE. This path encodes the multiband light curve into a latent representation and reconstructs the light curves. Using the partial light curve as input and the full light curve as the reconstruction target, the unsupervised path learns the latent representation of the data and establishes the connection between the partial light curve and the full light curve. The reconstruction process also enables the network to learn the underlying patterns that govern the data without any label information. In the supervised path, we add a fully connected layer to the shared weighted encoder to produce a classification probability. The supervised classification path forces the network to classify the light curve using label information.

\begin{figure}[!htb]
  \centering
  \includegraphics[width=1\textwidth]{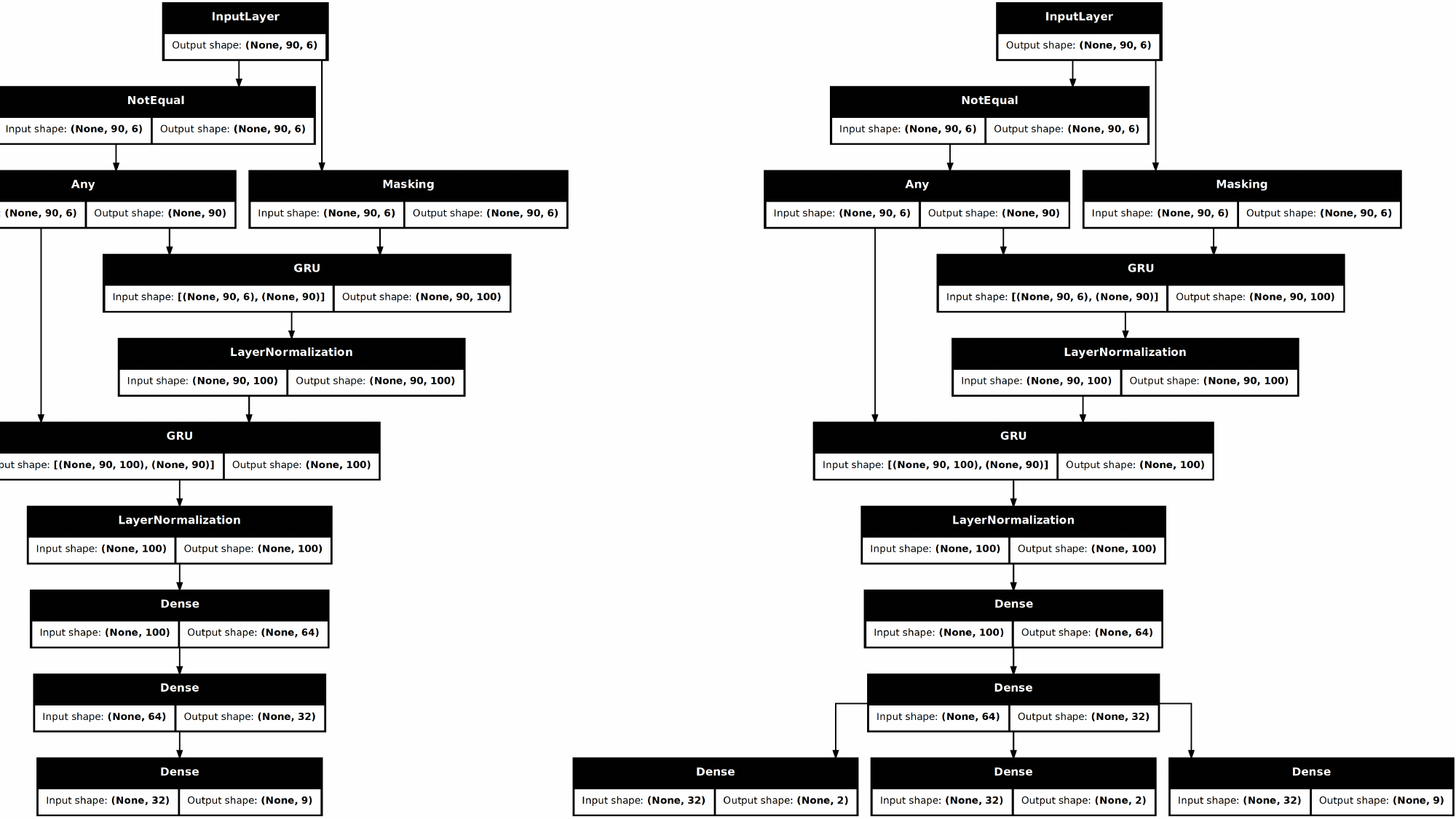}
  \caption{The detailed structure of the benchmark model (left panel) and the encoder module of our framework (right panel).}
  \label{fig:figure-6}
\end{figure}

\begin{figure}[!htb]
  \centering
  \includegraphics[width=0.3\textwidth]{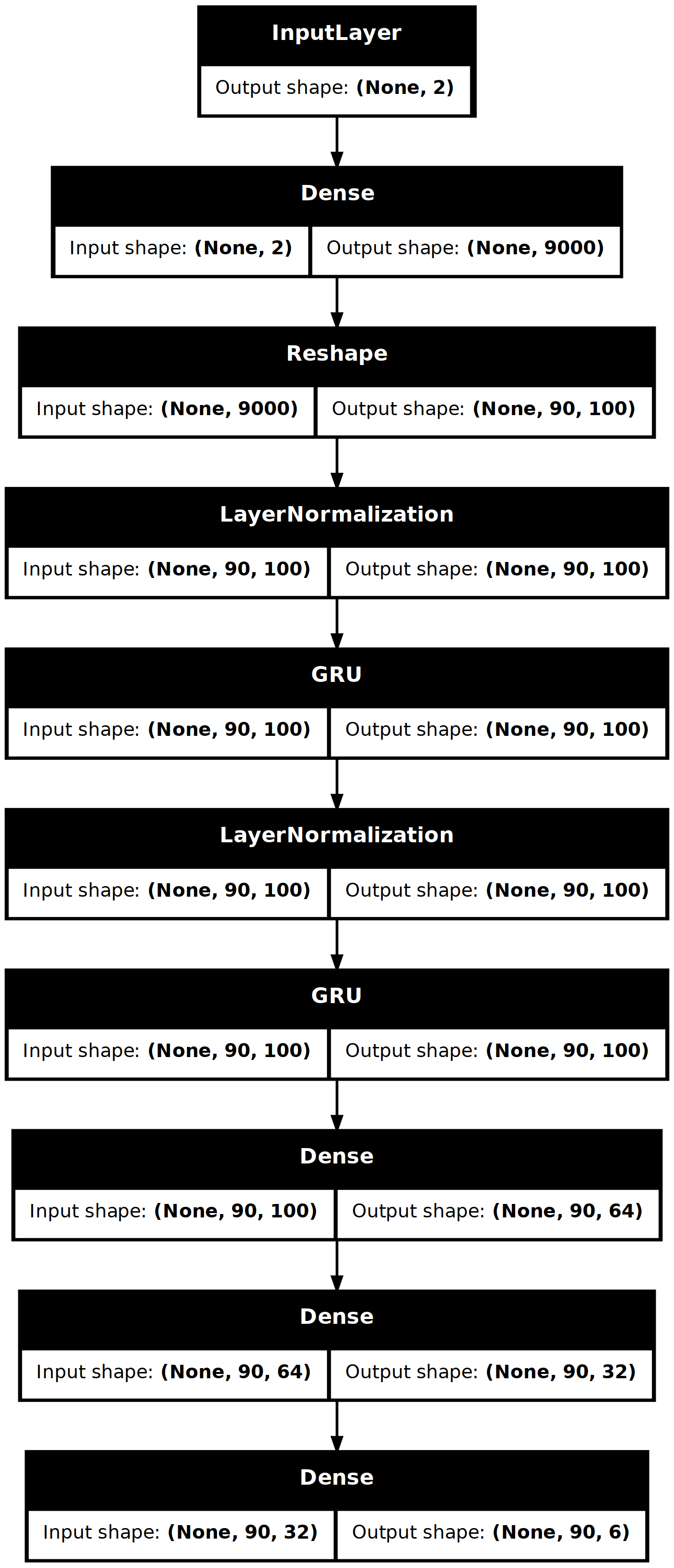}
  \caption{Decoder module of our framework}
  \label{fig:figure-7}
\end{figure}

We set the total loss function ($L_{\mathrm{tot}}$) of our framework as Equation ~\ref{equ:equation-6}. The total loss function is a weighted sum of three components. The KL loss ($L_{\mathrm{KL}}$, equation~\ref{equ:equation-4}) is the Kullback–Leibler divergence, which measures the difference between the probability distributions of the latent features and the Gaussian distribution. The loss of reconstruction ($L_{\mathrm{rec}}$, equation~\ref{equ:equation-3}) is the mean square error, which quantifies the difference between the reconstructed and the original data. Adding the Kullback-Leibler divergence ($D_{\mathrm{KL}}$) helps the VAE model obtain a better-structured latent space, allowing the auto-encoder to really understand the underlying patterns of the training data rather than simply replicating the original data. Due to the semi-supervised classification path and its classification task, we incorporate the categorical cross-entropy classification loss ($L_{\mathrm{cla}}$, equation~\ref{equ:equation-5}) into the final loss function to measure the deviation between the predicted labels and the true labels of each class. Given the different scales of these three losses,  we assign different weights $W_r=1, W_k=1, W_c=10$ to balance each component during the initial training steps.
\begin{equation}
    L_{\mathrm{rec}} = \sum_{i,j,t} (\hat{F}_{\mathrm{ij}}(t) - \tilde{F}_{\mathrm{ij}}(t))^2/N
    \label{equ:equation-3}
\end{equation}

\begin{equation}
    L_{\mathrm{KL}} = D_{KL}(P||Q)=\sum P\log(\frac{P}{Q})
    \label{equ:equation-4}
\end{equation}

\begin{equation}
    L_{\mathrm{cla}} = -\frac{1}{N}\sum^N_{i=1}\sum_{k=1}^Ky^i_klog(\hat y^i_k)
    \label{equ:equation-5}
\end{equation}

\begin{equation}
    L_{\mathrm{tot}} = W_r\times L_{\mathrm{rec}} + W_k \times L_{\mathrm{KL}} + W_c \times L_{\mathrm{cla}}
    \label{equ:equation-6}
\end{equation}
In these loss functions, $\hat{F}_{ij}(t)$ represents the reconstructed value of the light curve, $\tilde{F}_{ij}(t)$ represents the original light curve (after preprocessing); $N$ is the total number of light curve samples; $P$ and $Q$ are the probability distributions of the latent features and the Gaussian distribution, respectively; $K=9$ is the number of transient classes; and $y^i_k,\ \hat{y}^i_k$ are the true label and predicted label for class $k$, respectively.

A simple grid search approach is utilized to assess the effects of altering essential hyper parameters, including the number and size of hidden layers, based on established values from similar models. The hyper parameters selected are presented in the model topology diagrams~\ref{fig:figure-6} and~\ref{fig:figure-7}. These parameters were selected for their effectiveness in balancing adequate model capacity with overfitting prevention, meeting the demands of our current demonstration framework. Future tests with real-world data are expected to further refine these choices. During the training phase, we employed various \texttt{Keras} functions, such as metrics monitoring and adaptive learning rate adjustments, to ensure robust and reliable model training.

In the training phase, our proposed model functions as a framework with multiple inputs and outputs. Each branch has its own loss function, which is calculated separately and then combined as outlined in Equation~\ref{equ:equation-5} to produce the overall loss. This combined loss is used to optimize and update the model parameters. Once new samples are available, we update the model by first selecting the collected unlabeled data and eliminating any apparent artifacts. These cleaned unlabeled data are then utilized to continue training the framework together with the existing labeled data, ensuring the preservation of valuable information from prior data. During the inference phase, the framework processes real-time partial light curves as inputs and outputs the predicted complete light curve, as well as the predicted label probabilities.

\section{Results and Discussion} 
\label{sec:results and discussion}

In this section, we present the results of our work. First, we evaluate the multi-classification performance achieved using the supervised path. The test procedure is conducted through 10-fold cross-validation (each fold contains approximately 2000 light curves, randomly selected from the test dataset without specific division). This approach ensures robust evaluation by minimizing potential biases and providing a comprehensive assessment of model performance. Additionally, we compare our model's classification performance with that of a recurrent neural network (RNN) model, referred to hereafter as the benchmark model, using our simulated dataset. The architecture of the benchmark model closely mirrors the supervised path in our framework. Similar RNN structures have been used in previous studies, such as the early classification of transient light curves, and have demonstrated strong performance \citep{muthukrishna2019rapid, moller2020supernnova, becker2020scalable, gagliano2023first}. Therefore, the benchmark model serves as an appropriate point of comparison. Lastly, we briefly discuss the light curve reconstruction and latent space representation achieved through the unsupervised path, to assess the effectiveness of unsupervised learning.

\subsection{Multi-classification Performance}

\begin{figure}[!htb]
  \centering
  \includegraphics[width=1\textwidth]{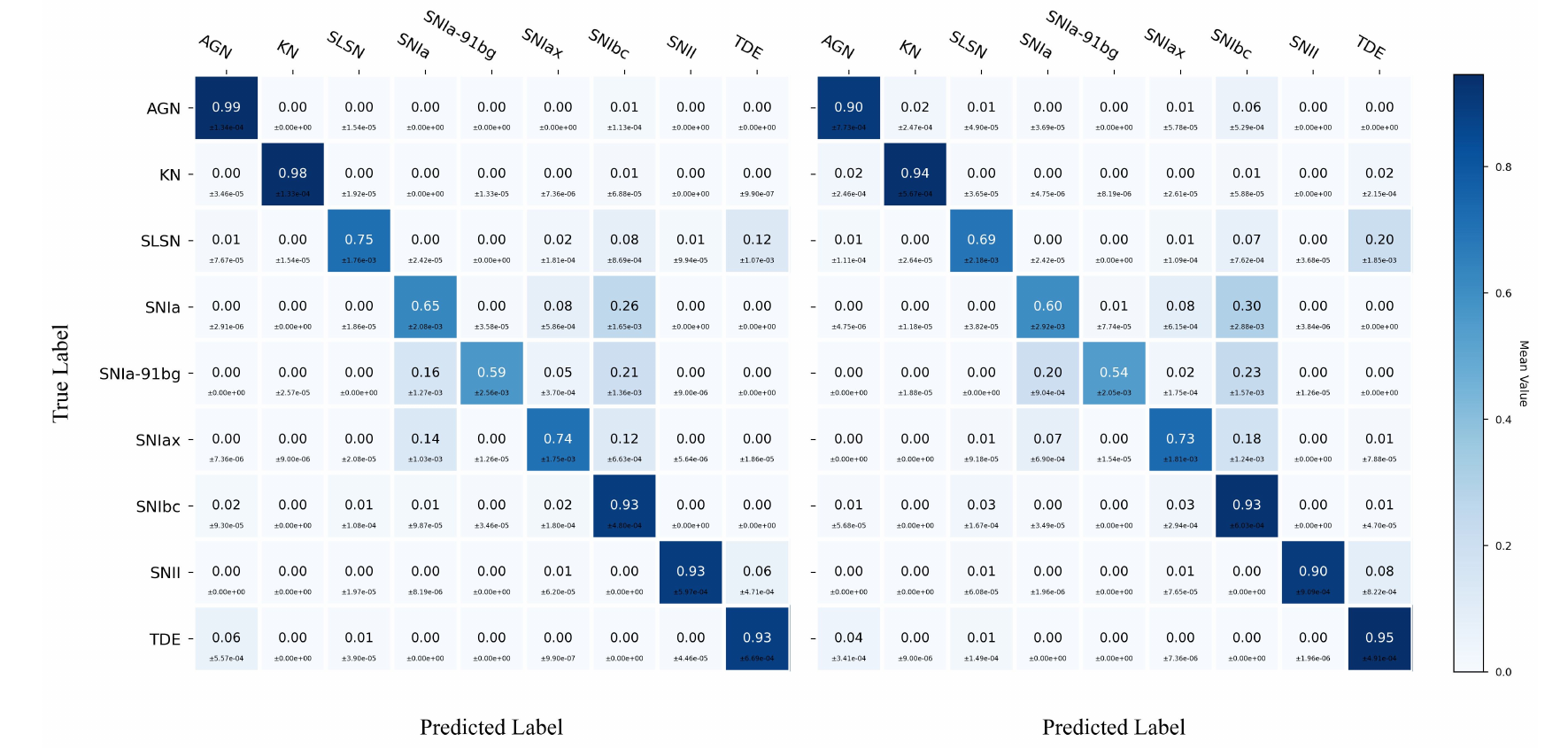}
  \caption{Comparison of confusion matrices (normalized over true labels) between the benchmark model (right panel) and our framework (left panel).}
  \label{fig:figure-8}
\end{figure}

We first evaluate the framework's classification performance using the confusion matrix. The confusion matrix directly reflects the model's classification performance in each category and can also be used to calculate various performance indicators. In the normalized confusion matrix, row i and column j represent the ratio in which the model classified instances of class i as class j based on the highest predicted probability. A better classifier will have most of its elements concentrated on the diagonal. In Figure ~\ref{fig:figure-8}, we plot the normalized confusion matrix of our framework compared to the benchmark model. We also calculate the corresponding precision and recall of each class based on the confusion matrix, as shown in Tables ~\ref{tab:table_2} and ~\ref{tab:table_3}. 

\startlongtable
\movetableright=-1.3in
\begin{table}
\caption{Precision comparison}
    \centering
    \begin{tabular}{cccccccccc}
       precision & AGN & KN & SLSN & SNIa & SNIa-91bg & SNIax & SNIbc & SNII & TDE \\
      \toprule
      our framework & 0.918 & \textbf{0.997} & \textbf{0.962} & \textbf{0.678} & 0.980 & 0.801 & \textbf{0.571} & 0.980 & \textbf{0.835}\\
      (uncertainty) & 4.8E-04 & 3.0E-05 & 5.3E-04 & 1.2E-03 & 2.3E-04 & 1.5E-03 & 7.9E-04 & 1.8E-04 & 8.7E-04\\
      \midrule
      benchmark model &0.918 & 0.967 & 0.906 & 0.677 & \textbf{0.981} & \textbf{0.808} & 0.524 & \textbf{0.994} & 0.747 \\
      (uncertainty) & 6.1E-04 & 3.05E-04 & 1.2E-03 & 1.5E-03 & 3.0E-04 & 1.3E-03 & 4.8E-04 & 6.6E-05 & 1.0E-03\\
      \bottomrule
    \end{tabular}

    \label{tab:table_2}
\end{table}

\startlongtable
\movetableright=-1.3in
\begin{table}
\caption{Recall comparison}
    \centering
    \begin{tabular}{cccccccccc}
       recall & AGN & KN & SLSN & SNIa & SNIa-91bg & SNIax & SNIbc & SNII & TDE \\
      \toprule
      our framework & \textbf{0.983} & \textbf{0.981} & \textbf{0.751} & \textbf{0.642} & \textbf{0.583} & \textbf{0.741} & \textbf{0.932} & \textbf{0.925} & 0.931\\
      (uncertainty) & 1.6E-04 & 1.7E-04 & 1.5E-03 & 2.0E-03 & 2.3E-03 & 2.0E-03 & 4.6E-04 & 6.9E-04 & 8.3E-04\\
      \midrule
      benchmark model & 0.904 & 0.937 & 0.688 & 0.602 & 0.541 & 0.729 & 0.926 & 0.900 & \textbf{0.945} \\
       (uncertainty) & 7.7E-04 & 5.6E-04 & 2.1E-03 & 2.9E-03 & 2.0E-03 & 1.8E-03 & 6.0E-04 & 9.0E-04 & 4.9E-04 \\
      \bottomrule
    \end{tabular}
    \label{tab:table_3}
\end{table}

\startlongtable
\movetableright=-1.3in
\begin{table}
\caption{Other performance comparison}
    \centering
    \begin{tabular}{cccccccccc}
       &macro P & macro R & macro F1 & accuracy & accuracy uncertainty \\
      \toprule
      our framework & \textbf{0.858} & \textbf{0.830} & \textbf{0.844} & \textbf{0.831} & 0.012 \\
      \midrule
      benchmark model &0.836 & 0.787 & 0.816 & 0.787 & 0.012 \\
      \bottomrule
    \end{tabular}

    \label{tab:table_4}
\end{table}

\begin{figure}[!htb]
  \centering
  \includegraphics[width=1\textwidth]{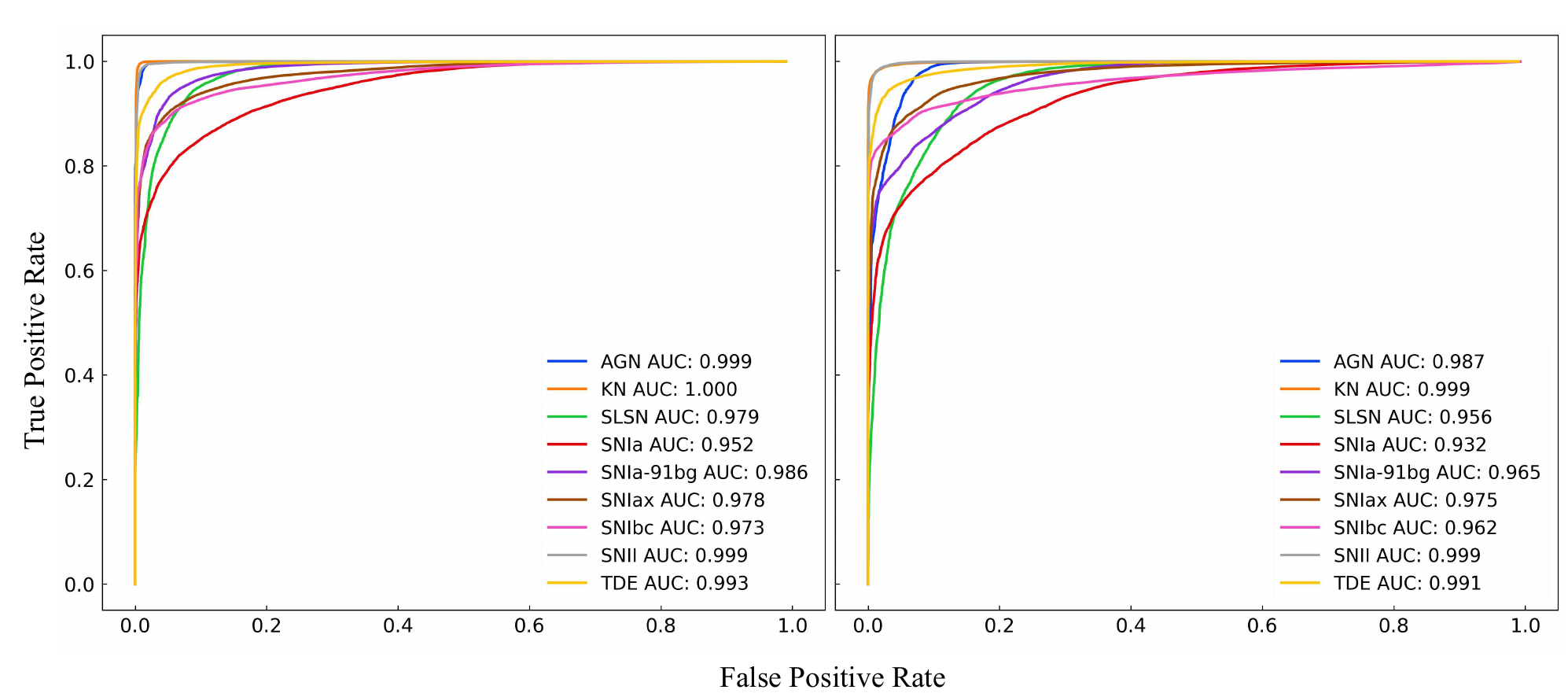}
  \caption{Comparison of ROC curve and corresponding AUC between the benchmark model (right panel) and our framework (left panel).}
  \label{fig:figure-9}
\end{figure}

\begin{figure}[!htb]
  \centering
  \includegraphics[width=0.6\textwidth]{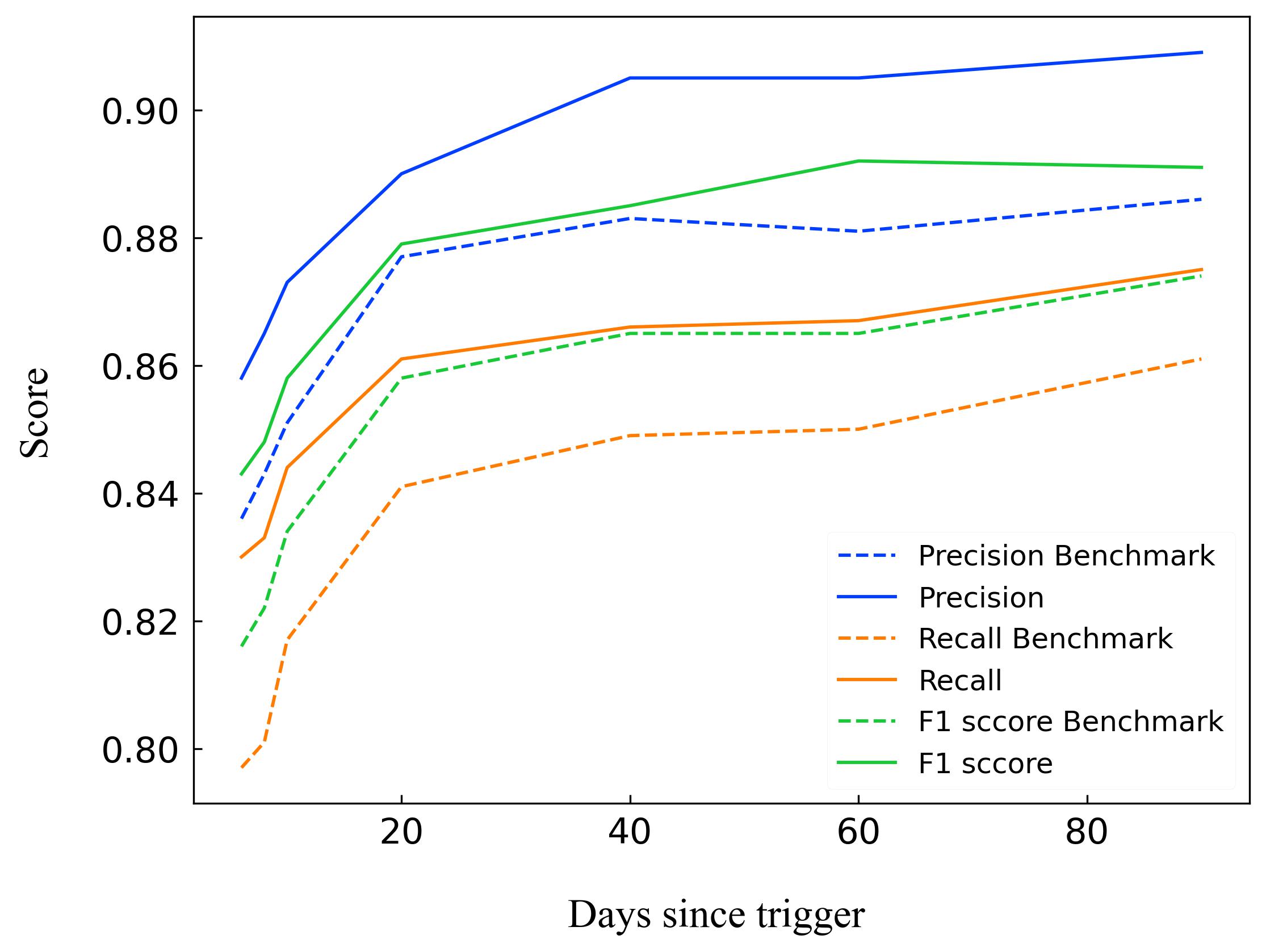}
  \caption{The macro precision, recall, and F1 scores of our framework are compared with those of the benchmark model across different phases of the light curve.}
  \label{fig:figure-10}
\end{figure}

We further examine the Receiver Operating Characteristics (ROC) curve of our framework. Compared to the confusion matrix, the ROC curve and Area Under the Curve (AUC) provide a more detailed view of the classifier's performance. The classifier outputs a probability distribution rather than a simple true or false prediction. This probability information can help to gain deeper insight into the behavior of the classifier. To construct the ROC curve for each class, we calculate the true positive rate (TPR) and the false positive rate (FPR) at various thresholds. In a multi-classification scenario, TPR is the ratio of correctly classified objects in a specific class to the total number of objects in that class. Meanwhile, the FPR is the ratio of incorrectly classified objects in all other classes to the total number of objects in those classes. Figure~\ref{fig:figure-9} shows the ROC curve and AUC for both our framework (right) and the benchmark model (left).
Overall, our framework outperforms the benchmark model in the classification task. Both the confusion matrix and the ROC curve visually demonstrate improved classification performance in all subclasses. The promising performance in the early phase is auspicious for our ability to identify transient candidates soon after detection, allowing us to pursue critical follow-up resources effectively. Meanwhile, Figure~\ref{fig:figure-10} shows that our framework performs consistently over time. 

\subsection{Latent Representation}
\begin{figure}[!htb]
  \centering
    \includegraphics[width=1\textwidth]{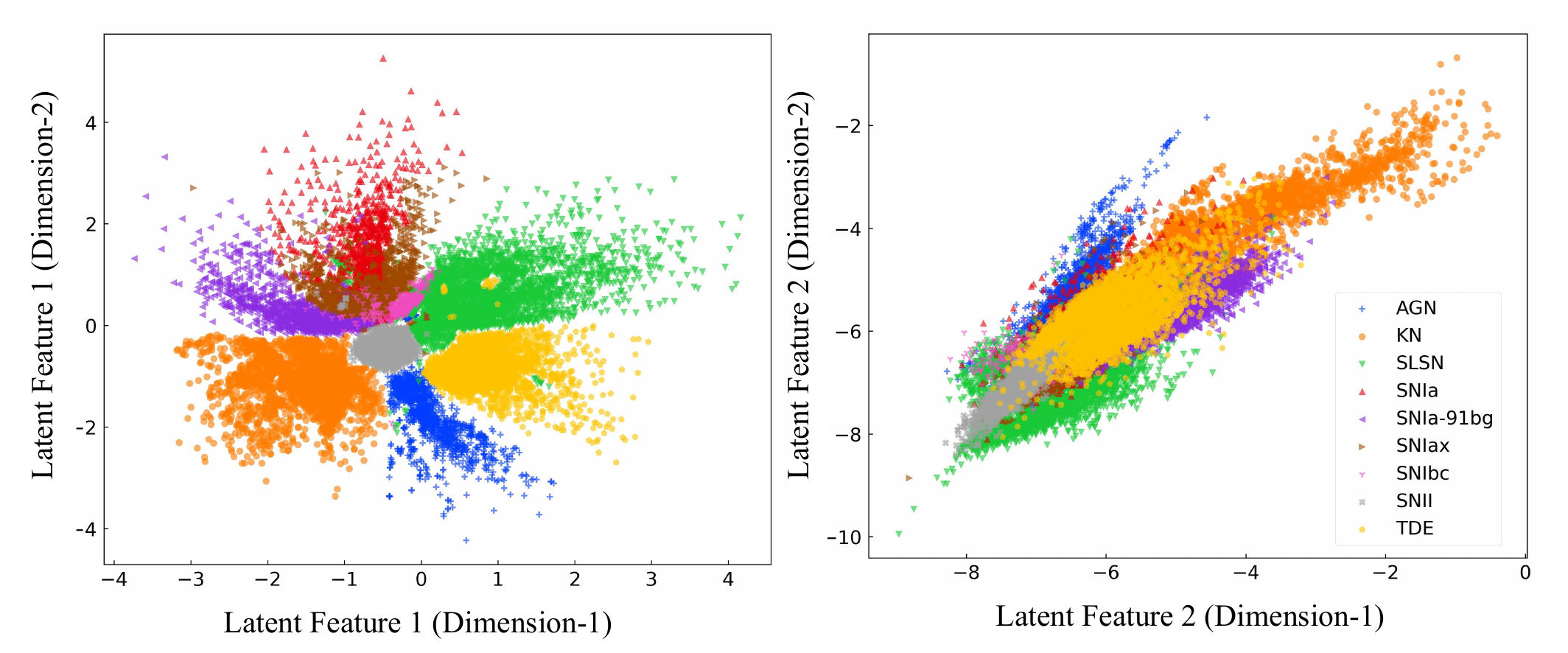}
  \caption{The latent space representation of the early light curve in the test dataset. The two axes are the latent features learned by the unsupervised path.}
  \label{fig:figure-11}
\end{figure}

\begin{figure}[!htb]
  \centering
  \includegraphics[width=1\textwidth]{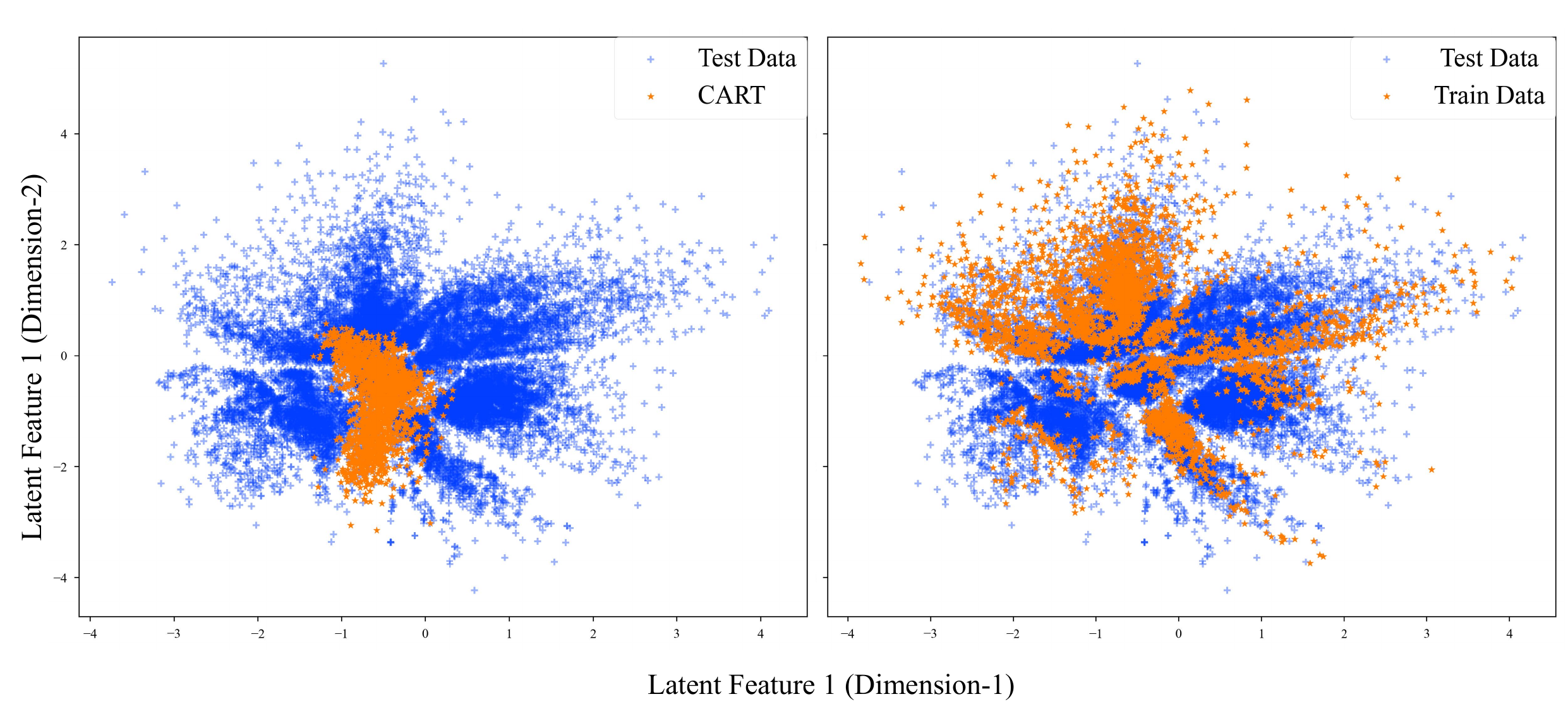}
  \caption{Latent space distribution. The left panel is the different distribution of the Cart and test data set; The right panel is the different distribution of the train dataset and the test dataset.}
  \label{fig:figure-12}
\end{figure}

To further investigate the latent space, we plot a two-dimensional latent representation of the data, which has been optimized for classification tasks using a supervised learning approach. As shown in Figure~\ref{fig:figure-11}, the latent features of the test data clearly exhibit clustering characteristics. This clustering is particularly pronounced in the first latent feature, leading us to focus on this feature space. Different classes display varying degrees of dispersion in the latent space, which may indicate the inherent variability within each class. We also compare the latent representations of the training and test data. As illustrated in Figure~\ref{fig:figure-12}, the training data occupy a smaller portion of the latent space compared to the test data. This visualization allows for a clear inspection of the potential bias in the training data, as it highlights the regions of the latent space that are underrepresented in the training set.

Finally, we exclude the CART (Calcium-rich gap transients) class in the training dataset  \citep[as they are relatively rare, see ][]{kessler2019models} and visualize the distribution of the latent representation of this unseen category in our framework. The left panel in Figure~\ref{fig:figure-12} shows some overlap between CART and the test dataset, as well as some distinct differences. In summary, the latent space provides an interesting perspective on the underlying patterns that drive different transient light curve behaviors. As the amount of unlabeled data increases, this latent space will be further refined, making the results more reliable.

\subsection{Light Curve Reconstruction}

Reconstruction of the light curve plays a crucial role in learning the underlying low-dimensional distributions. The right panel in Figure~\ref{fig:figure-13} shows some examples of a reconstructed full light curve compared to the original light curve. The average reconstructed loss is 19.56\% (measured by the mean square reconstruction loss, as defined in Equation~\ref{equ:equation-3}) in the test data set. The results indicate that the model can effectively reconstruct the original light curve given the latent representation of the unlabeled data. In other words, the unlabeled data can help the encoder module learn the underlying patterns of different sources, specifically in terms of reconstructing the photometric light curve. Visual inspection of the reconstructed light curve demonstrates the ability of the unsupervised path, even without the label information. Through this reconstruction, the framework also establishes a connection between the early light curve and its later evolution.

The reconstructed light curve also provides possibilities for further analysis, such as forecasting future observation values and anomaly detection. One can compare the predicted light curve with actual observations \citep{muthukrishna2022real} or utilize the latent representation of the light curve \citep{villar2020anomaly}. The reconstructed light curve and its corresponding latent representation provide a unique perspective for human experts to investigate the unlabeled dataset. The latent space is currently optimized for classification, but it can be further adjusted for other purposes, which requires further investigation and is likely a promising target for future research.

4\subsection{Discussion}

The above results not only validate our original intentions but also highlight the novelty of our framework. A key feature of our framework is its optimization for the early classification task. By incorporating an unsupervised auto-encoder path, our classifier takes the early light curve as input and reconstructs the full light curve. This unique approach allows the model to leverage the information carried by the full light curve when processing the partial light curve, thereby enhancing the performance of the classification path. Another significant feature of our framework is its ability to make full use of massive unlabeled data. The framework can learn the latent representation of unlabeled data while maintaining its capability to classify different objects using label information. This is particularly valuable for a new survey telescope. The number of spectrally identified objects is limited, and the scarcity of labeled astronomical objects often constrains the effectiveness of supervised machine learning algorithms for classification. At the same time, even considering the observational characteristics in the simulation, differences could occur when applying the simulation results to real observation. Thus, our framework provides a valuable tool for WFST and future wide-field survey telescopes.

\begin{figure}[!htb]
  \centering
  \includegraphics[width=1\textwidth]{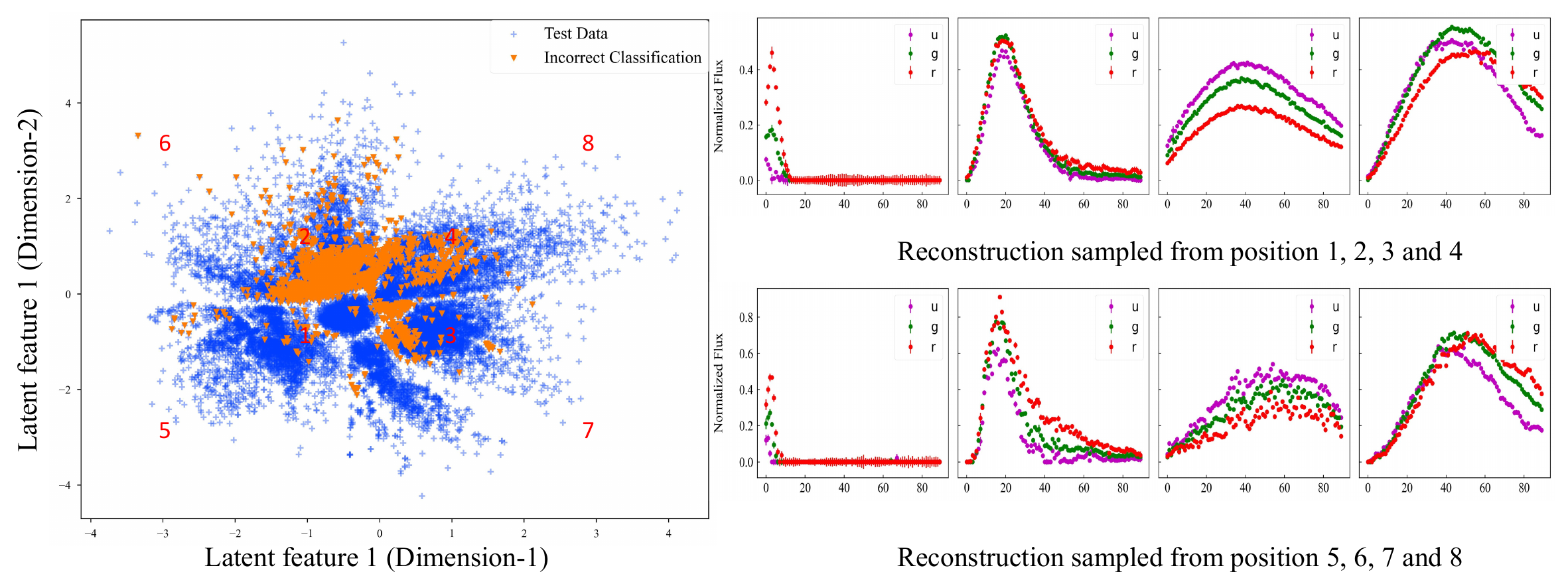}
  \caption{The latent space analysis reveals distinct patterns in model performance: the left panel highlights misclassified data points within the context of the full test set, while the right panel visualizes reconstructed light curves sampled from diverse regions of the latent space, demonstrating its structural heterogeneity.}
  \label{fig:figure-13}
\end{figure}

It is also worth mentioning some shortcomings of our method and potential improvements that we plan to carry out in future work. 
First, we have observed that the GP can sometimes generate suboptimal interpolations, particularly in cases where there are large observational gaps. Despite these limitations, the reconstruction path in our framework can still learn a general pattern. Hypothesize that this discrepancy arises because the GP is optimized for individual light curves, whereas our framework is optimized based on the entire training data set. This highlights the importance of integrating machine learning methods to complement GP.  Research on more effective ways to handle missing values while maintaining optimal performance would be beneficial.

Another consideration is that models based on simulated data often exhibit overly optimistic performance when applied to real-world data, likely due to biases in the simulation. \textbf{As observed in our results, there are no imposed differences in class distribution between the training and test sets, and the split between different datasets is based solely on the redshift. These conditions may deviate from real-world settings. This also explains the small uncertainty in Table \ref{tab:table_2} and Table \ref{tab:table_3}}. In this paper, we focus on demonstrating the relative performance of the unsupervised approach compared to a regular architecture, since this comparison is less affected by the differences between real and simulated data. However, any classification tool must ultimately serve real observations. To this end, we plan to further refine our framework using real data from the WFST.

Finally, we analyze the characteristics of objects incorrectly classified within our framework. As shown in Figure \ref{fig:figure-13}, the latent distribution of these objects reveals two primary scenarios where misclassification tends to occur: first, SNIa and their variants (such as SNIax, SNIa-91bg, and SNIbc) are often misclassified despite having a small reconstruction error, suggesting that the internal similarities between these objects make them difficult to distinguish, a finding supported by the structure of the latent space; second, misclassification is more frequent when the reconstruction error is large, indicating that our framework may require further refinement, particularly when dealing with out-of-distribution samples. Figure \ref{fig:figure-13} also illustrates reconstructed light curves from various regions of the latent space, showing that samples closer to the test data are better reconstructed, while those farther away exhibit poorer reconstruction quality. However, even in poorly reconstructed light curves, some basic shapes can still be discerned.

\section{Summary}

In summary, we introduce a semi-supervised classification framework tailored for the Wide Field Survey Telescope. This framework is grounded in well-established research and integrates several novel enhancements. First, prior research has demonstrated the effectiveness of unsupervised auto-encoder architecture in harnessing information from vast amounts of unlabeled data. We adapt this architecture for the task of early classification. Second, we incorporate a supervised classification route, training both pathways concurrently. This approach ensures that our framework retains its classification abilities while also learning latent representations from new unlabeled data.

Throughout the experiments, our framework exhibits the following benefits:
\begin{itemize}
    \item Its semi-supervised nature allows for effective updates with new unlabeled data.
    \item These characteristics significantly improve the classification results compared to traditional supervised classification approaches, including the benchmark model used in our work. 
    \item Our framework improves precision and recall across each subclass (Section~\ref{sec:results and discussion}).
    \item Our framework provides latent representations of light curves and reconstructed data, which can be incorporated into other techniques, offering flexibility and adaptability to diverse downstream applications, like anomaly detection and the determination of the observation phase.
\end{itemize}

Currently, our system is utilized for multi-classification tasks to recognize interesting transients shortly after WFST discovers them. We plan to explore further applications. Furthermore, our objective is to improve target selection and anomaly detection using real observational data from WFST, focusing on selecting TDEs or identifying rare objects, which are of significant interest to WFST.

\section*{Acknowledgements}
We would like to express our sincere gratitude to the anonymous referee for their valuable and insightful comments. The Wide Field Survey Telescope (WFST) is a joint facility of the University of Science and Technology of China, Purple Mountain Observatory. This work is supported by National Key Research and Development Program of China (2023YFA1608100). LF gratefully acknowledges the support of the National Natural Science Foundation of China (NSFC, grant No. 12173037, 12233008), the CAS Project for Young Scientists in Basic Research (No. YSBR-092), the Fundamental Research Funds for the Central Universities (WK3440000006) and Cyrus Chun Ying Tang Foundations.

\vspace{5mm}

\facilities{WFST: 2.5 m}

\software{\texttt{keras} \citep{chollet2015keras}, \texttt{george} \citep{2015ITPAM..38..252A} , \texttt{SNANA} \citep{kessler2009snana}, \texttt{scikit-learn} \citep{scikit-learn}}

\bibliography{ms}

\begin{thebibliography}{}
\expandafter\ifx\csname natexlab\endcsname\relax\def\natexlab#1{#1}\fi
\providecommand{\url}[1]{\href{#1}{#1}}
\providecommand{\dodoi}[1]{doi:~\href{http://doi.org/#1}{\nolinkurl{#1}}}
\providecommand{\doeprint}[1]{\href{http://ascl.net/#1}{\nolinkurl{http://ascl.net/#1}}}
\providecommand{\doarXiv}[1]{\href{https://arxiv.org/abs/#1}{\nolinkurl{https://arxiv.org/abs/#1}}}

\bibitem[{{Abdelhadi} \& {Rubin}(2024)}]{abdelhadi2024picture}
{Abdelhadi}, B., \& {Rubin}, D. 2024, \pasp, 136, 124504,
  \dodoi{10.1088/1538-3873/ad9a7d}

\bibitem[{{Aguirre} {et~al.}(2019){Aguirre}, {Pichara}, \&
  {Becker}}]{aguirre2019deep}
{Aguirre}, C., {Pichara}, K., \& {Becker}, I. 2019, \mnras, 482, 5078,
  \dodoi{10.1093/mnras/sty2836}

\bibitem[{{Allam} \& {McEwen}(2024)}]{allam2024paying}
{Allam}, T., \& {McEwen}, J.~D. 2024, RAS Techniques and Instruments, 3, 209,
  \dodoi{10.1093/rasti/rzad046}

\bibitem[{{Allam} {et~al.}(2023){Allam}, {Peloton}, \&
  {McEwen}}]{allam2023tiny}
{Allam}, Jr., T., {Peloton}, J., \& {McEwen}, J.~D. 2023, arXiv e-prints,
  arXiv:2303.08951, \dodoi{10.48550/arXiv.2303.08951}

\bibitem[{{Ambikasaran} {et~al.}(2015){Ambikasaran}, {Foreman-Mackey},
  {Greengard}, {Hogg}, \& {O'Neil}}]{2015ITPAM..38..252A}
{Ambikasaran}, S., {Foreman-Mackey}, D., {Greengard}, L., {Hogg}, D.~W., \&
  {O'Neil}, M. 2015, IEEE Transactions on Pattern Analysis and Machine
  Intelligence, 38, 252, \dodoi{10.1109/TPAMI.2015.2448083}

\bibitem[{{Baron}(2019)}]{baron2019machine}
{Baron}, D. 2019, arXiv e-prints, arXiv:1904.07248,
  \dodoi{10.48550/arXiv.1904.07248}

\bibitem[{{Becker} {et~al.}(2020){Becker}, {Pichara}, {Catelan}, {Protopapas},
  {Aguirre}, \& {Nikzat}}]{becker2020scalable}
{Becker}, I., {Pichara}, K., {Catelan}, M., {et~al.} 2020, \mnras, 493, 2981,
  \dodoi{10.1093/mnras/staa350}

\bibitem[{{Bellm} {et~al.}(2019){Bellm}, {Kulkarni}, {Graham}, {Dekany},
  {Smith}, {Riddle}, {Masci}, {Helou}, {Prince}, {Adams}, {Barbarino},
  {Barlow}, {Bauer}, {Beck}, {Belicki}, {Biswas}, {Blagorodnova}, {Bodewits},
  {Bolin}, {Brinnel}, {Brooke}, {Bue}, {Bulla}, {Burruss}, {Cenko}, {Chang},
  {Connolly}, {Coughlin}, {Cromer}, {Cunningham}, {De}, {Delacroix}, {Desai},
  {Duev}, {Eadie}, {Farnham}, {Feeney}, {Feindt}, {Flynn}, {Franckowiak},
  {Frederick}, {Fremling}, {Gal-Yam}, {Gezari}, {Giomi}, {Goldstein},
  {Golkhou}, {Goobar}, {Groom}, {Hacopians}, {Hale}, {Henning}, {Ho}, {Hover},
  {Howell}, {Hung}, {Huppenkothen}, {Imel}, {Ip}, {Ivezi{\'c}}, {Jackson},
  {Jones}, {Juric}, {Kasliwal}, {Kaspi}, {Kaye}, {Kelley}, {Kowalski},
  {Kramer}, {Kupfer}, {Landry}, {Laher}, {Lee}, {Lin}, {Lin}, {Lunnan},
  {Giomi}, {Mahabal}, {Mao}, {Miller}, {Monkewitz}, {Murphy}, {Ngeow},
  {Nordin}, {Nugent}, {Ofek}, {Patterson}, {Penprase}, {Porter}, {Rauch},
  {Rebbapragada}, {Reiley}, {Rigault}, {Rodriguez}, {van Roestel}, {Rusholme},
  {van Santen}, {Schulze}, {Shupe}, {Singer}, {Soumagnac}, {Stein}, {Surace},
  {Sollerman}, {Szkody}, {Taddia}, {Terek}, {Van Sistine}, {van Velzen},
  {Vestrand}, {Walters}, {Ward}, {Ye}, {Yu}, {Yan}, \&
  {Zolkower}}]{bellm2018zwicky}
{Bellm}, E.~C., {Kulkarni}, S.~R., {Graham}, M.~J., {et~al.} 2019, \pasp, 131,
  018002, \dodoi{10.1088/1538-3873/aaecbe}

\bibitem[{{Bond} {et~al.}(2001){Bond}, {Abe}, {Dodd}, {Hearnshaw}, {Honda},
  {Jugaku}, {Kilmartin}, {Marles}, {Masuda}, {Matsubara}, {Muraki}, {Nakamura},
  {Nankivell}, {Noda}, {Noguchi}, {Ohnishi}, {Rattenbury}, {Reid}, {Saito},
  {Sato}, {Sekiguchi}, {Skuljan}, {Sullivan}, {Sumi}, {Takeuti}, {Watase},
  {Wilkinson}, {Yamada}, {Yanagisawa}, \& {Yock}}]{bond2001real}
{Bond}, I.~A., {Abe}, F., {Dodd}, R.~J., {et~al.} 2001, \mnras, 327, 868,
  \dodoi{10.1046/j.1365-8711.2001.04776.x}

\bibitem[{{Boone}(2019)}]{boone2019avocado}
{Boone}, K. 2019, \aj, 158, 257, \dodoi{10.3847/1538-3881/ab5182}

\bibitem[{{Boone}(2021)}]{boone2021parsnip}
---. 2021, \aj, 162, 275, \dodoi{10.3847/1538-3881/ac2a2d}

\bibitem[{{Burhanudin} \& {Maund}(2023)}]{burhanudin2023pan}
{Burhanudin}, U.~F., \& {Maund}, J.~R. 2023, \mnras, 521, 1601,
  \dodoi{10.1093/mnras/stac3672}

\bibitem[{{Cai} {et~al.}(2025){Cai}, {Xu}, {Fan}, {Wan}, {Kong}, {Hu}, {Jiang},
  {Hu}, {Zhu}, {Li}, {Lin}, {Fang}, {Xue}, {Zhen}, \& {Wang}}]{cai2025WFST}
{Cai}, M., {Xu}, Z., {Fan}, L., {et~al.} 2025, arXiv e-prints,
  arXiv:2501.15018, \dodoi{10.48550/arXiv.2501.15018}

\bibitem[{{Chaini} \& {Kumar}(2020)}]{chaini2020astronomical}
{Chaini}, S., \& {Kumar}, S.~S. 2020, arXiv e-prints, arXiv:2006.12333,
  \dodoi{10.48550/arXiv.2006.12333}

\bibitem[{{Cho} {et~al.}(2014){Cho}, {van Merrienboer}, {Bahdanau}, \&
  {Bengio}}]{cho2014properties}
{Cho}, K., {van Merrienboer}, B., {Bahdanau}, D., \& {Bengio}, Y. 2014, arXiv
  e-prints, arXiv:1409.1259, \dodoi{10.48550/arXiv.1409.1259}

\bibitem[{Chollet {et~al.}(2015)}]{chollet2015keras}
Chollet, F., {et~al.} 2015, Keras, \url{https://keras.io}

\bibitem[{{Dark Energy Survey Collaboration} {et~al.}(2016){Dark Energy Survey
  Collaboration}, {Abbott}, {Abdalla}, {Aleksi{\'c}}, {Allam}, {Amara},
  {Bacon}, {Balbinot}, {Banerji}, {Bechtol}, {Benoit-L{\'e}vy}, {Bernstein},
  {Bertin}, {Blazek}, {Bonnett}, {Bridle}, {Brooks}, {Brunner}, {Buckley-Geer},
  {Burke}, {Caminha}, {Capozzi}, {Carlsen}, {Carnero-Rosell}, {Carollo},
  {Carrasco-Kind}, {Carretero}, {Castander}, {Clerkin}, {Collett}, {Conselice},
  {Crocce}, {Cunha}, {D'Andrea}, {da Costa}, {Davis}, {Desai}, {Diehl},
  {Dietrich}, {Dodelson}, {Doel}, {Drlica-Wagner}, {Estrada}, {Etherington},
  {Evrard}, {Fabbri}, {Finley}, {Flaugher}, {Foley}, {Fosalba}, {Frieman},
  {Garc{\'\i}a-Bellido}, {Gaztanaga}, {Gerdes}, {Giannantonio}, {Goldstein},
  {Gruen}, {Gruendl}, {Guarnieri}, {Gutierrez}, {Hartley}, {Honscheid}, {Jain},
  {James}, {Jeltema}, {Jouvel}, {Kessler}, {King}, {Kirk}, {Kron}, {Kuehn},
  {Kuropatkin}, {Lahav}, {Li}, {Lima}, {Lin}, {Maia}, {Makler}, {Manera},
  {Maraston}, {Marshall}, {Martini}, {McMahon}, {Melchior}, {Merson}, {Miller},
  {Miquel}, {Mohr}, {Morice-Atkinson}, {Naidoo}, {Neilsen}, {Nichol}, {Nord},
  {Ogando}, {Ostrovski}, {Palmese}, {Papadopoulos}, {Peiris}, {Peoples},
  {Percival}, {Plazas}, {Reed}, {Refregier}, {Romer}, {Roodman}, {Ross},
  {Rozo}, {Rykoff}, {Sadeh}, {Sako}, {S{\'a}nchez}, {Sanchez}, {Santiago},
  {Scarpine}, {Schubnell}, {Sevilla-Noarbe}, {Sheldon}, {Smith}, {Smith},
  {Soares-Santos}, {Sobreira}, {Soumagnac}, {Suchyta}, {Sullivan}, {Swanson},
  {Tarle}, {Thaler}, {Thomas}, {Thomas}, {Tucker}, {Vieira}, {Vikram},
  {Walker}, {Wechsler}, {Weller}, {Wester}, {Whiteway}, {Wilcox}, {Yanny},
  {Zhang}, \& {Zuntz}}]{dark2016dark}
{Dark Energy Survey Collaboration}, {Abbott}, T., {Abdalla}, F.~B., {et~al.}
  2016, \mnras, 460, 1270, \dodoi{10.1093/mnras/stw641}

\bibitem[{{Donoso-Oliva} {et~al.}(2023){Donoso-Oliva}, {Becker}, {Protopapas},
  {Cabrera-Vives}, {Vishnu}, \& {Vardhan}}]{donoso2023astromer}
{Donoso-Oliva}, C., {Becker}, I., {Protopapas}, P., {et~al.} 2023, \aap, 670,
  A54, \dodoi{10.1051/0004-6361/202243928}

\bibitem[{{Drake} {et~al.}(2012){Drake}, {Djorgovski}, {Mahabal}, {Prieto},
  {Beshore}, {Graham}, {Catalan}, {Larson}, {Christensen}, {Donalek}, \&
  {Williams}}]{djorgovski2011catalina}
{Drake}, A.~J., {Djorgovski}, S.~G., {Mahabal}, A., {et~al.} 2012, in IAU
  Symposium, Vol. 285, New Horizons in Time Domain Astronomy, ed. E.~{Griffin},
  R.~{Hanisch}, \& R.~{Seaman}, 306--308, \dodoi{10.1017/S1743921312000889}

\bibitem[{{Frontera-Pons} {et~al.}(2017){Frontera-Pons}, {Sureau}, {Bobin}, \&
  {Le Floc'h}}]{frontera2017unsupervised}
{Frontera-Pons}, J., {Sureau}, F., {Bobin}, J., \& {Le Floc'h}, E. 2017, \aap,
  603, A60, \dodoi{10.1051/0004-6361/201630240}

\bibitem[{{Gagliano} {et~al.}(2023){Gagliano}, {Contardo}, {Foreman-Mackey},
  {Malz}, \& {Aleo}}]{gagliano2023first}
{Gagliano}, A., {Contardo}, G., {Foreman-Mackey}, D., {Malz}, A.~I., \& {Aleo},
  P.~D. 2023, \apj, 954, 6, \dodoi{10.3847/1538-4357/ace326}

\bibitem[{{Hlo{\v{z}}ek} {et~al.}(2023){Hlo{\v{z}}ek}, {Malz}, {Ponder}, {Dai},
  {Narayan}, {Ishida}, {Allam}, {Bahmanyar}, {Bi}, {Biswas}, {Boone}, {Chen},
  {Du}, {Erdem}, {Galbany}, {Garreta}, {Jha}, {Jones}, {Kessler}, {Lin}, {Liu},
  {Lochner}, {Mahabal}, {Mandel}, {Margolis}, {Mart{\'\i}nez-Galarza},
  {McEwen}, {Muthukrishna}, {Nakatsuka}, {Noumi}, {Oya}, {Peiris}, {Peters},
  {Puget}, {Setzer}, {Siddhartha}, {Stefanov}, {Xie}, {Yan}, {Yeh}, \&
  {Zuo}}]{hlovzek2023results}
{Hlo{\v{z}}ek}, R., {Malz}, A.~I., {Ponder}, K.~A., {et~al.} 2023, \apjs, 267,
  25, \dodoi{10.3847/1538-4365/accd6a}

\bibitem[{Hochreiter \& Schmidhuber(1997)}]{hochreiter1997long}
Hochreiter, S., \& Schmidhuber, J. 1997, Neural computation, 9, 1735

\bibitem[{{Hosenie} {et~al.}(2020){Hosenie}, {Lyon}, {Stappers}, {Mootoovaloo},
  \& {McBride}}]{hosenie2020imbalance}
{Hosenie}, Z., {Lyon}, R., {Stappers}, B., {Mootoovaloo}, A., \& {McBride}, V.
  2020, \mnras, 493, 6050, \dodoi{10.1093/mnras/staa642}

\bibitem[{{Hu} {et~al.}(2022{\natexlab{a}}){Hu}, {Wang}, {Chen}, \&
  {Yang}}]{hu2022image}
{Hu}, L., {Wang}, L., {Chen}, X., \& {Yang}, J. 2022{\natexlab{a}}, \apj, 936,
  157, \dodoi{10.3847/1538-4357/ac7394}

\bibitem[{{Hu} {et~al.}(2022{\natexlab{b}}){Hu}, {Hu}, {Jiang}, {Xiao}, {Fan},
  {Wei}, \& {Wu}}]{Humaokai_2023}
{Hu}, M., {Hu}, L., {Jiang}, J.-a., {et~al.} 2022{\natexlab{b}}, Universe, 9,
  7, \dodoi{10.3390/universe9010007}

\bibitem[{{Ivezi{\'c}} {et~al.}(2019){Ivezi{\'c}}, {Kahn}, {Tyson}, {Abel},
  {Acosta}, {Allsman}, {Alonso}, {AlSayyad}, {Anderson}, {Andrew}, {Angel},
  {Angeli}, {Ansari}, {Antilogus}, {Araujo}, {Armstrong}, {Arndt}, {Astier},
  {Aubourg}, {Auza}, {Axelrod}, {Bard}, {Barr}, {Barrau}, {Bartlett}, {Bauer},
  {Bauman}, {Baumont}, {Bechtol}, {Bechtol}, {Becker}, {Becla}, {Beldica},
  {Bellavia}, {Bianco}, {Biswas}, {Blanc}, {Blazek}, {Blandford}, {Bloom},
  {Bogart}, {Bond}, {Booth}, {Borgland}, {Borne}, {Bosch}, {Boutigny},
  {Brackett}, {Bradshaw}, {Brandt}, {Brown}, {Bullock}, {Burchat}, {Burke},
  {Cagnoli}, {Calabrese}, {Callahan}, {Callen}, {Carlin}, {Carlson},
  {Chandrasekharan}, {Charles-Emerson}, {Chesley}, {Cheu}, {Chiang}, {Chiang},
  {Chirino}, {Chow}, {Ciardi}, {Claver}, {Cohen-Tanugi}, {Cockrum}, {Coles},
  {Connolly}, {Cook}, {Cooray}, {Covey}, {Cribbs}, {Cui}, {Cutri}, {Daly},
  {Daniel}, {Daruich}, {Daubard}, {Daues}, {Dawson}, {Delgado}, {Dellapenna},
  {de Peyster}, {de Val-Borro}, {Digel}, {Doherty}, {Dubois},
  {Dubois-Felsmann}, {Durech}, {Economou}, {Eifler}, {Eracleous}, {Emmons},
  {Fausti Neto}, {Ferguson}, {Figueroa}, {Fisher-Levine}, {Focke}, {Foss},
  {Frank}, {Freemon}, {Gangler}, {Gawiser}, {Geary}, {Gee}, {Geha}, {Gessner},
  {Gibson}, {Gilmore}, {Glanzman}, {Glick}, {Goldina}, {Goldstein}, {Goodenow},
  {Graham}, {Gressler}, {Gris}, {Guy}, {Guyonnet}, {Haller}, {Harris},
  {Hascall}, {Haupt}, {Hernandez}, {Herrmann}, {Hileman}, {Hoblitt}, {Hodgson},
  {Hogan}, {Howard}, {Huang}, {Huffer}, {Ingraham}, {Innes}, {Jacoby}, {Jain},
  {Jammes}, {Jee}, {Jenness}, {Jernigan}, {Jevremovi{\'c}}, {Johns}, {Johnson},
  {Johnson}, {Jones}, {Juramy-Gilles}, {Juri{\'c}}, {Kalirai}, {Kallivayalil},
  {Kalmbach}, {Kantor}, {Karst}, {Kasliwal}, {Kelly}, {Kessler}, {Kinnison},
  {Kirkby}, {Knox}, {Kotov}, {Krabbendam}, {Krughoff}, {Kub{\'a}nek},
  {Kuczewski}, {Kulkarni}, {Ku}, {Kurita}, {Lage}, {Lambert}, {Lange},
  {Langton}, {Le Guillou}, {Levine}, {Liang}, {Lim}, {Lintott}, {Long},
  {Lopez}, {Lotz}, {Lupton}, {Lust}, {MacArthur}, {Mahabal}, {Mandelbaum},
  {Markiewicz}, {Marsh}, {Marshall}, {Marshall}, {May}, {McKercher}, {McQueen},
  {Meyers}, {Migliore}, {Miller}, \& {Mills}}]{ivezic2019lsst}
{Ivezi{\'c}}, {\v{Z}}., {Kahn}, S.~M., {Tyson}, J.~A., {et~al.} 2019, \apj,
  873, 111, \dodoi{10.3847/1538-4357/ab042c}

\bibitem[{{Jamal} \& {Bloom}(2020)}]{jamal2020neural}
{Jamal}, S., \& {Bloom}, J.~S. 2020, \apjs, 250, 30,
  \dodoi{10.3847/1538-4365/aba8ff}

\bibitem[{{Jedicke} {et~al.}(2012){Jedicke}, {Tonry}, {Veres}, {Farnocchia},
  {Spoto}, {Rest}, {Wainscoat}, \& {Lee}}]{jedicke2012atlas}
{Jedicke}, R., {Tonry}, J., {Veres}, P., {et~al.} 2012, in AAS/Division for
  Planetary Sciences Meeting Abstracts, Vol.~44, AAS/Division for Planetary
  Sciences Meeting Abstracts \#44, 210.12

\bibitem[{{Kaiser} {et~al.}(2010){Kaiser}, {Burgett}, {Chambers}, {Denneau},
  {Heasley}, {Jedicke}, {Magnier}, {Morgan}, {Onaka}, \&
  {Tonry}}]{kaiser2010pan}
{Kaiser}, N., {Burgett}, W., {Chambers}, K., {et~al.} 2010, in Society of
  Photo-Optical Instrumentation Engineers (SPIE) Conference Series, Vol. 7733,
  Ground-based and Airborne Telescopes III, ed. L.~M. {Stepp}, R.~{Gilmozzi},
  \& H.~J. {Hall}, 77330E, \dodoi{10.1117/12.859188}

\bibitem[{{Kessler} {et~al.}(2009){Kessler}, {Bernstein}, {Cinabro}, {Dilday},
  {Frieman}, {Jha}, {Kuhlmann}, {Miknaitis}, {Sako}, {Taylor}, \&
  {Vanderplas}}]{kessler2009snana}
{Kessler}, R., {Bernstein}, J.~P., {Cinabro}, D., {et~al.} 2009, \pasp, 121,
  1028, \dodoi{10.1086/605984}

\bibitem[{{Kessler} {et~al.}(2019){Kessler}, {Narayan}, {Avelino}, {Bachelet},
  {Biswas}, {Brown}, {Chernoff}, {Connolly}, {Dai}, {Daniel}, {Di Stefano},
  {Drout}, {Galbany}, {Gonz{\'a}lez-Gait{\'a}n}, {Graham}, {Hlo{\v{z}}ek},
  {Ishida}, {Guillochon}, {Jha}, {Jones}, {Mandel}, {Muthukrishna}, {O'Grady},
  {Peters}, {Pierel}, {Ponder}, {Pr{\v{s}}a}, {Rodney}, {Villar}, {LSST Dark
  Energy Science Collaboration}, \& {Transient and Variable Stars Science
  Collaboration}}]{kessler2019models}
{Kessler}, R., {Narayan}, G., {Avelino}, A., {et~al.} 2019, \pasp, 131, 094501,
  \dodoi{10.1088/1538-3873/ab26f1}

\bibitem[{{Lei} {et~al.}(2023){Lei}, {Zhu}, {Kong}, {Wang}, {Zheng}, {Shi},
  {Fan}, \& {Liu}}]{Lei_2023}
{Lei}, L., {Zhu}, Q.-F., {Kong}, X., {et~al.} 2023, Research in Astronomy and
  Astrophysics, 23, 035013, \dodoi{10.1088/1674-4527/acb877}

\bibitem[{{Lin} {et~al.}(2022){Lin}, {Jiang}, \& {Kong}}]{LinZheyu_2022}
{Lin}, Z., {Jiang}, N., \& {Kong}, X. 2022, \mnras, 513, 2422,
  \dodoi{10.1093/mnras/stac946}

\bibitem[{{Liu} {et~al.}(2023){Liu}, {Lin}, {Yu}, {Wang}, {Mourani}, {Zhao}, \&
  {Dai}}]{Liuzhengyan_2023}
{Liu}, Z.-Y., {Lin}, Z.-Y., {Yu}, J.-M., {et~al.} 2023, \apj, 947, 59,
  \dodoi{10.3847/1538-4357/acc73b}

\bibitem[{{Ma} {et~al.}(2018){Ma}, {Zhu}, {Li}, \& {Xu}}]{ma2018radio}
{Ma}, Z., {Zhu}, J., {Li}, W., \& {Xu}, H. 2018, arXiv e-prints,
  arXiv:1806.00398, \dodoi{10.48550/arXiv.1806.00398}

\bibitem[{Modelers(2022)}]{plasticc_modelers_2022_6672739}
Modelers, P. 2022, {Libraries \& Recommended Citations for using PLAsTiCC
  Models}, v4,  Zenodo, \dodoi{10.5281/zenodo.6672739}

\bibitem[{{M{\"o}ller} \& {de Boissi{\`e}re}(2020)}]{moller2020supernnova}
{M{\"o}ller}, A., \& {de Boissi{\`e}re}, T. 2020, \mnras, 491, 4277,
  \dodoi{10.1093/mnras/stz3312}

\bibitem[{{Muthukrishna} {et~al.}(2022){Muthukrishna}, {Mandel}, {Lochner},
  {Webb}, \& {Narayan}}]{muthukrishna2022real}
{Muthukrishna}, D., {Mandel}, K.~S., {Lochner}, M., {Webb}, S., \& {Narayan},
  G. 2022, \mnras, 517, 393, \dodoi{10.1093/mnras/stac2582}

\bibitem[{{Muthukrishna} {et~al.}(2019){Muthukrishna}, {Narayan}, {Mandel},
  {Biswas}, \& {Hlo{\v{z}}ek}}]{muthukrishna2019rapid}
{Muthukrishna}, D., {Narayan}, G., {Mandel}, K.~S., {Biswas}, R., \&
  {Hlo{\v{z}}ek}, R. 2019, \pasp, 131, 118002, \dodoi{10.1088/1538-3873/ab1609}

\bibitem[{{Narayan} {et~al.}(2018){Narayan}, {Zaidi}, {Soraisam}, \& {ANTARES
  Collaboration}}]{narayan2018machine}
{Narayan}, G., {Zaidi}, T., {Soraisam}, M., \& {ANTARES Collaboration}. 2018,
  in American Astronomical Society Meeting Abstracts, Vol. 231, American
  Astronomical Society Meeting Abstracts \#231, 332.08

\bibitem[{{Naul} {et~al.}(2018){Naul}, {Bloom}, {P{\'e}rez}, \& {van der
  Walt}}]{naul2018recurrent}
{Naul}, B., {Bloom}, J.~S., {P{\'e}rez}, F., \& {van der Walt}, S. 2018, Nature
  Astronomy, 2, 151, \dodoi{10.1038/s41550-017-0321-z}

\bibitem[{{Pasquet} {et~al.}(2019){Pasquet}, {Pasquet}, {Chaumont}, \&
  {Fouchez}}]{pasquet2019pelican}
{Pasquet}, J., {Pasquet}, J., {Chaumont}, M., \& {Fouchez}, D. 2019, \aap, 627,
  A21, \dodoi{10.1051/0004-6361/201834473}

\bibitem[{{Pedregosa} {et~al.}(2011){Pedregosa}, {Varoquaux}, {Gramfort},
  {Michel}, {Thirion}, {Grisel}, {Blondel}, {M{\"u}ller}, {Nothman}, {Louppe},
  {Prettenhofer}, {Weiss}, {Dubourg}, {Vanderplas}, {Passos}, {Cournapeau},
  {Brucher}, {Perrot}, \& {Duchesnay}}]{scikit-learn}
{Pedregosa}, F., {Varoquaux}, G., {Gramfort}, A., {et~al.} 2011, Journal of
  Machine Learning Research, 12, 2825, \dodoi{10.48550/arXiv.1201.0490}

\bibitem[{{Pimentel} {et~al.}(2023){Pimentel}, {Est{\'e}vez}, \&
  {F{\"o}rster}}]{pimentel2022deep}
{Pimentel}, {\'O}., {Est{\'e}vez}, P.~A., \& {F{\"o}rster}, F. 2023, \aj, 165,
  18, \dodoi{10.3847/1538-3881/ac9ab4}

\bibitem[{{Portillo} {et~al.}(2020){Portillo}, {Parejko}, {Vergara}, \&
  {Connolly}}]{portillo2020dimensionality}
{Portillo}, S. K.~N., {Parejko}, J.~K., {Vergara}, J.~R., \& {Connolly}, A.~J.
  2020, \aj, 160, 45, \dodoi{10.3847/1538-3881/ab9644}

\bibitem[{{Pursiainen} {et~al.}(2018){Pursiainen}, {Childress}, {Smith},
  {Prajs}, {Sullivan}, {Davis}, {Foley}, {Asorey}, {Calcino}, {Carollo},
  {Curtin}, {D'Andrea}, {Glazebrook}, {Gutierrez}, {Hinton}, {Hoormann},
  {Inserra}, {Kessler}, {King}, {Kuehn}, {Lewis}, {Lidman}, {Macaulay},
  {M{\"o}ller}, {Nichol}, {Sako}, {Sommer}, {Swann}, {Tucker}, {Uddin},
  {Wiseman}, {Zhang}, {Abbott}, {Abdalla}, {Allam}, {Annis}, {Avila}, {Brooks},
  {Buckley-Geer}, {Burke}, {Carnero Rosell}, {Carrasco Kind}, {Carretero},
  {Castander}, {Cunha}, {Davis}, {De Vicente}, {Diehl}, {Doel}, {Eifler},
  {Flaugher}, {Fosalba}, {Frieman}, {Garc{\'\i}a-Bellido}, {Gruen}, {Gruendl},
  {Gutierrez}, {Hartley}, {Hollowood}, {Honscheid}, {James}, {Jeltema},
  {Kuropatkin}, {Li}, {Lima}, {Maia}, {Martini}, {Menanteau}, {Ogando},
  {Plazas}, {Roodman}, {Sanchez}, {Scarpine}, {Schindler}, {Smith},
  {Soares-Santos}, {Sobreira}, {Suchyta}, {Swanson}, {Tarle}, {Tucker},
  {Walker}, \& {DES Collaboration}}]{pursiainen2018rapidly}
{Pursiainen}, M., {Childress}, M., {Smith}, M., {et~al.} 2018, \mnras, 481,
  894, \dodoi{10.1093/mnras/sty2309}

\bibitem[{{Qu} \& {Sako}(2022)}]{qu2022photometric}
{Qu}, H., \& {Sako}, M. 2022, \aj, 163, 57, \dodoi{10.3847/1538-3881/ac39a1}

\bibitem[{{Reis} {et~al.}(2019){Reis}, {Baron}, \&
  {Shahaf}}]{reis2018probabilistic}
{Reis}, I., {Baron}, D., \& {Shahaf}, S. 2019, \aj, 157, 16,
  \dodoi{10.3847/1538-3881/aaf101}

\bibitem[{{Revsbech} {et~al.}(2018){Revsbech}, {Trotta}, \& {van
  Dyk}}]{revsbech2018staccato}
{Revsbech}, E.~A., {Trotta}, R., \& {van Dyk}, D.~A. 2018, \mnras, 473, 3969,
  \dodoi{10.1093/mnras/stx2570}

\bibitem[{{Richards} {et~al.}(2012){Richards}, {Homrighausen}, {Freeman},
  {Schafer}, \& {Poznanski}}]{richards2012semi}
{Richards}, J.~W., {Homrighausen}, D., {Freeman}, P.~E., {Schafer}, C.~M., \&
  {Poznanski}, D. 2012, \mnras, 419, 1121,
  \dodoi{10.1111/j.1365-2966.2011.19768.x}

\bibitem[{{Riess} {et~al.}(1998){Riess}, {Filippenko}, {Challis},
  {Clocchiatti}, {Diercks}, {Garnavich}, {Gilliland}, {Hogan}, {Jha},
  {Kirshner}, {Leibundgut}, {Phillips}, {Reiss}, {Schmidt}, {Schommer},
  {Smith}, {Spyromilio}, {Stubbs}, {Suntzeff}, \&
  {Tonry}}]{riess1998observational}
{Riess}, A.~G., {Filippenko}, A.~V., {Challis}, P., {et~al.} 1998, \aj, 116,
  1009, \dodoi{10.1086/300499}

\bibitem[{{Russeil} {et~al.}(2024){Russeil}, {Malanchev}, {Aleo}, {Ishida},
  {Pruzhinskaya}, {Gangler}, {Lavrukhina}, {Volnova}, {Voloshina},
  {Semenikhin}, {Sreejith}, {Kornilov}, \& {Korolev}}]{russeil2024rainbow}
{Russeil}, E., {Malanchev}, K.~L., {Aleo}, P.~D., {et~al.} 2024, \aap, 683,
  A251, \dodoi{10.1051/0004-6361/202348158}

\bibitem[{{Sako} {et~al.}(2008){Sako}, {Bassett}, {Becker}, {Cinabro},
  {DeJongh}, {Depoy}, {Dilday}, {Doi}, {Frieman}, {Garnavich}, {Hogan},
  {Holtzman}, {Jha}, {Kessler}, {Konishi}, {Lampeitl}, {Marriner}, {Miknaitis},
  {Nichol}, {Prieto}, {Riess}, {Richmond}, {Romani}, {Schneider}, {Smith},
  {SubbaRao}, {Takanashi}, {Tokita}, {van der Heyden}, {Yasuda}, {Zheng},
  {Barentine}, {Brewington}, {Choi}, {Dembicky}, {Harnavek}, {Ihara}, {Im},
  {Ketzeback}, {Kleinman}, {Krzesi{\'n}ski}, {Long}, {Malanushenko},
  {Malanushenko}, {McMillan}, {Morokuma}, {Nitta}, {Pan}, {Saurage}, \&
  {Snedden}}]{sako2007sloan}
{Sako}, M., {Bassett}, B., {Becker}, A., {et~al.} 2008, \aj, 135, 348,
  \dodoi{10.1088/0004-6256/135/1/348}

\bibitem[{{S{\'a}nchez-S{\'a}ez} {et~al.}(2021){S{\'a}nchez-S{\'a}ez}, {Reyes},
  {Valenzuela}, {F{\"o}rster}, {Eyheramendy}, {Elorrieta}, {Bauer},
  {Cabrera-Vives}, {Est{\'e}vez}, {Catelan}, {Pignata}, {Huijse}, {De Cicco},
  {Ar{\'e}valo}, {Carrasco-Davis}, {Abril}, {Kurtev}, {Borissova}, {Arredondo},
  {Castillo-Navarrete}, {Rodriguez}, {Ruz-Mieres}, {Moya},
  {Sabatini-Gacit{\'u}a}, {Sep{\'u}lveda-Cobo}, \&
  {Camacho-I{\~n}iguez}}]{sanchez2021alert}
{S{\'a}nchez-S{\'a}ez}, P., {Reyes}, I., {Valenzuela}, C., {et~al.} 2021, \aj,
  161, 141, \dodoi{10.3847/1538-3881/abd5c1}

\bibitem[{{Su} {et~al.}(2024){Su}, {Cai}, {Wang}, {Wang}, {Xue}, {Cai}, {Fan},
  {Guo}, {He}, {He}, {Hu}, {Jiang}, {Jiang}, {Kang}, {Lei}, {Liu}, {Liu},
  {Liu}, {Sheng}, {Sun}, \& {Zhao}}]{Su_2024}
{Su}, Z.-B., {Cai}, Z.-Y., {Wang}, J.-X., {et~al.} 2024, \apj, 976, 155,
  \dodoi{10.3847/1538-4357/ad86bc}

\bibitem[{{Tomaney} \& {Crotts}(1996)}]{tomaney1996expanding}
{Tomaney}, A.~B., \& {Crotts}, A. P.~S. 1996, \aj, 112, 2872,
  \dodoi{10.1086/118228}

\bibitem[{{Villar} {et~al.}(2020{\natexlab{a}}){Villar}, {Cranmer}, {Contardo},
  {Ho}, \& {Yao-Yu Lin}}]{villar2020anomaly}
{Villar}, V.~A., {Cranmer}, M., {Contardo}, G., {Ho}, S., \& {Yao-Yu Lin}, J.
  2020{\natexlab{a}}, arXiv e-prints, arXiv:2010.11194,
  \dodoi{10.48550/arXiv.2010.11194}

\bibitem[{{Villar} {et~al.}(2019){Villar}, {Nicholl}, \&
  {Berger}}]{villar2018superluminous}
{Villar}, V.~A., {Nicholl}, M., \& {Berger}, E. 2019, in American Astronomical
  Society Meeting Abstracts, Vol. 233, American Astronomical Society Meeting
  Abstracts \#233, 131.01

\bibitem[{{Villar} {et~al.}(2020{\natexlab{b}}){Villar}, {Hosseinzadeh},
  {Berger}, {Ntampaka}, {Jones}, {Challis}, {Chornock}, {Drout}, {Foley},
  {Kirshner}, {Lunnan}, {Margutti}, {Milisavljevic}, {Sanders}, {Pan}, {Rest},
  {Scolnic}, {Magnier}, {Metcalfe}, {Wainscoat}, \&
  {Waters}}]{villar2020superraenn}
{Villar}, V.~A., {Hosseinzadeh}, G., {Berger}, E., {et~al.} 2020{\natexlab{b}},
  \apj, 905, 94, \dodoi{10.3847/1538-4357/abc6fd}

\bibitem[{{Wang} {et~al.}(2023){Wang}, {Liu}, {Cai}, {Geng}, {Fang}, {He},
  {Jiang}, {Jiang}, {Kong}, {Li}, {Li}, {Luo}, {Pan}, {Wu}, {Yang}, {Yu},
  {Zheng}, {Zhu}, {Cai}, {Chen}, {Chen}, {Dai}, {Fan}, {Fan}, {Fang}, {He},
  {Hu}, {Hu}, {Jin}, {Jiang}, {Li}, {Li}, {Li}, {Liang}, {Lin}, {Liu}, {Liu},
  {Liu}, {Liu}, {Liu}, {Lou}, {Qu}, {Sheng}, {Shi}, {Shu}, {Su}, {Sun}, {Wang},
  {Wang}, {Wang}, {Wang}, {Wei}, {Wei}, {Xue}, {Yan}, {Yang}, {Yuan}, {Yuan},
  {Zhang}, {Zhang}, {Zhao}, \& {Zhao}}]{wang2023sciences}
{Wang}, T., {Liu}, G., {Cai}, Z., {et~al.} 2023, Science China Physics,
  Mechanics, and Astronomy, 66, 109512, \dodoi{10.1007/s11433-023-2197-5}

\bibitem[{Yu {et~al.}(2021)Yu, Li, Zhang, Xiao, Cui, Tao, Tang, Sun, \&
  Bi}]{yu2021survey}
Yu, C., Li, K., Zhang, Y., {et~al.} 2021, Wiley Interdisciplinary Reviews: Data
  Mining and Knowledge Discovery, 11, e1425

\end{thebibliography}
\bibliographystyle{aasjournal}
\end{CJK*}
\end{document}